\documentclass[pre,twocolumn,amsmath,amssymb]{revtex4b}

\usepackage{graphicx}
\usepackage{amssymb}
\usepackage{amsmath}
\usepackage{enumerate}

\begin{document}

\title{Cytoskeleton and Cell Motility}

\author{Thomas Risler}
\affiliation{
Institut Curie, Centre de Recherche, UMR 168\\
(UPMC Univ Paris 06, CNRS),\\
26 rue d'Ulm, F-75005 Paris, France
}

\maketitle

\section*{\label{Se.Outline}\large{Article Outline}}

\begin{description}
\item {\bf Glossary}
\item {\bf I. Definition of the Subject and Its Importance}
\item {\bf II. Introduction}
\item {\bf III. The Diversity of Cell Motility}
\begin{description}
\item A. Swimming
\item B. Crawling
\item C. Extensions of cell motility
\end{description}
\item {\bf IV. The Cell Cytoskeleton}
\begin{description}
\item A. Biopolymers
\item B. Molecular motors
\item C. Motor families
\item D. Other cytoskeleton-associated proteins
\item E. Cell anchoring and regulatory pathways
\item F. The prokaryotic cytoskeleton
\end{description}
\item {\bf V. Filament-Driven Motility}
\begin{description}
\item A. Microtubule growth and catastrophes
\item B. Actin gels
\item C. Modeling polymerization forces
\item D. A model system for studying actin-based motility: The bacterium \emph{Listeria monocytogenes}
\item E. Another example of filament-driven amoeboid motility: The nematode sperm cell
\end{description}
\item {\bf VI. Motor-Driven Motility}
\begin{description}
\item A. Generic considerations
\item B. Phenomenological description close to thermodynamic equilibrium
\item C. Hopping and transport models
\item D. The two-state model
\item E. Coupled motors and spontaneous oscillations
\item F. Axonemal beating
\end{description}
\item {\bf VII. Putting It Together: Active Polymer Solutions}
\begin{description}
\item A. Mesoscopic approaches
\item B. Microscopic approaches
\item C. Macroscopic phenomenological approaches: The active gels
\item D. Comparisons of the different approaches to describing active polymer solutions
\end{description}
\item {\bf VIII. Extensions and Future Directions}
\item {\bf Acknowledgments}
\item {\bf Bibliography}
\end{description}

\newpage

\section*{\label{Se.Glossary}\large{Glossary}}

\begin{description}
\item[Cell] Structural and functional elementary unit of all life forms. The cell is the smallest unit that can be characterized as living.
\item[Eukaryotic cell] Cell that possesses a nucleus, a small membrane-bounded compartment that contains the genetic material of the cell. Cells that lack a nucleus are called \emph{prokaryotic cells} or \emph{prokaryotes}.
\item[Domains of life] \emph{archaea},  \emph{bacteria} and \emph{eukarya} - or in English \emph{eukaryotes}, and made of eukaryotic cells - which constitute the three fundamental branches in which all life forms are classified. Archaea and bacteria are prokaryotes. All multicellular organisms are eukaryotes, but eukaryotes can also be single-cell organisms. Eukaryotes are usually classified into four kingdoms: animals, plants, fungi and protists.
\item[Motility] Spontaneous, self-generated movement of a biological system.
\item[Cytoskeleton] System of protein filaments crisscrossing the inner part of the cell and which, with the help of the many proteins that interact with it, enables the cell to insure its structural integrity and morphology, exert forces and produce motion.
\item[Amoeboid motility] Crawling locomotion of a eukaryotic cell by means of protrusion of its leading edge.
\item[Molecular Motor] Motor of molecular size. In this context, protein or macromolecular complex that converts a specific source of energy into mechanical work.
\item[Filament] Here, extended unidimensional structure made of an assembly of repeated protein units that hold together via physical interactions (without covalent bonds). A filament will be either a single \emph{polymer} (or here \emph{biopolymer}), a linear assembly of such polymers, or a linear assembly of molecular motors.
\item[Active gel] Cross-linked network of linear or branched polymers interacting by physical means, and that is dynamically driven out of equilibrium by a source of energy.
\end{description}

\newpage

\section{\label{Se.Motivation}Definition of the Subject and Its Importance}

We, as human beings, are made of a collection of cells, which are most commonly considered as the elementary building blocks of all living forms on earth \cite{cellB}. Whether they belong to each of the three domains of life (\emph{archaea}, \emph{bacteria} or \emph{eukarya}), cells are small membrane-bounded compartments that are capable of homeostasis, metabolism, response to their environment, growth, reproduction, adaptation through evolution and, at the cellular as well as multicellular level, organization. In addition, spontaneous, self-generated movement - also known as \emph{motility} - is one of the properties that we most closely associate with all life forms. Even in the case of apparently inanimate living forms on macroscopic scales, like most plants and fungi, constitutive cells are constantly remodeling their internal structure for the entire organism to perform its metabolism, growth and reproduction \cite{brayB}. In animals like human beings, cell motility is at the basis of most - if not all - essential processes participating in their lifetime, from their development, maintenance, to eventual death. It is indeed crucially involved for example in embryonic development (where individual as well as collective motions of cells underly morphogenesis), wound healing and recovery from injuries (where cellular migration is essential for tissue repair and regeneration), as well as immune response and most of disease progressions. There, on a biomedical point of view, cellular motility is involved in processes as diverse as neutrophils (white blood cells) and macrophages (cells that ingest bacteria) progressions, axonal regrowth after injuries, multiple sclerosis and cancer metastases. In addition, motility defects of the animal cells themselves can lead to a variety of inherited health problems, including male infertility, deafness and chronic inflammatory diseases.

\section{\label{Se.Intro}Introduction}

\begin{figure}[h]
\scalebox{0.1}{
\includegraphics{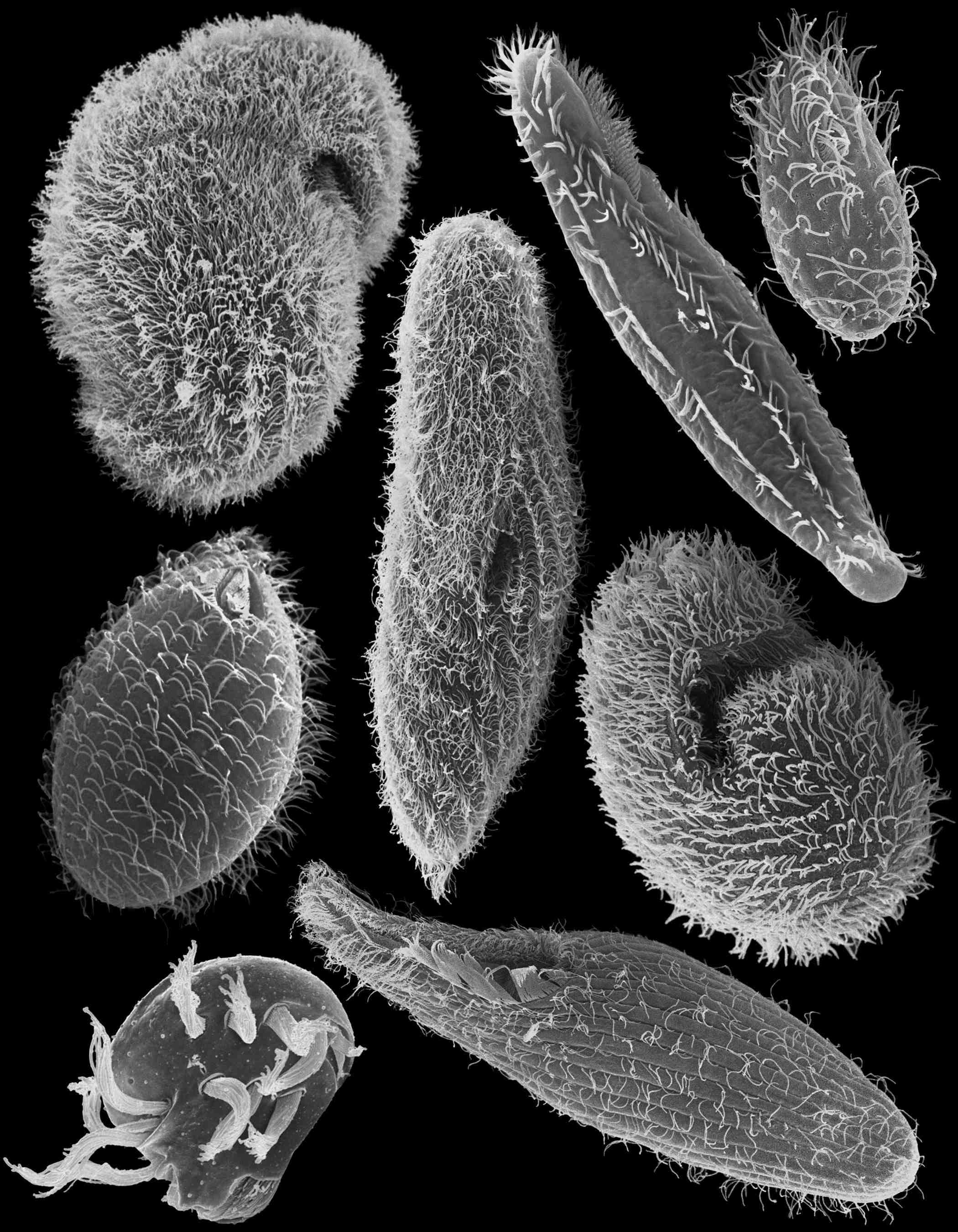}
}
\caption[cytoskeletonandcellmotility_fig1]
{\label{cytoskeletonandcellmotility_fig1}
Electron micrographs of different species of ciliated \emph{protozoa}. Almost all members of the protozoan group are non-pathogenic free-living organisms. Source: Foissner, W. and Zankl, A. (unpublished).}
\end{figure}

\begin{figure}[h]
\scalebox{0.22}{
\includegraphics{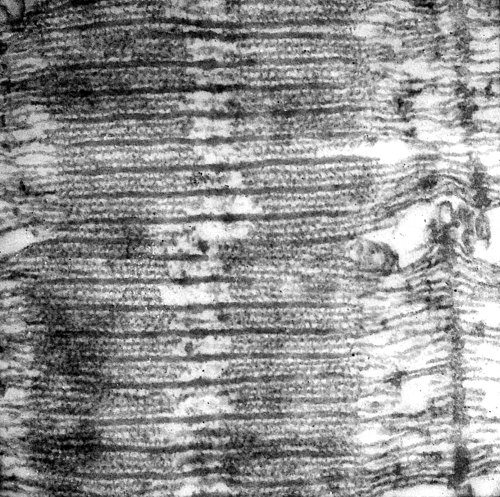}
}
\scalebox{0.22}{
\includegraphics{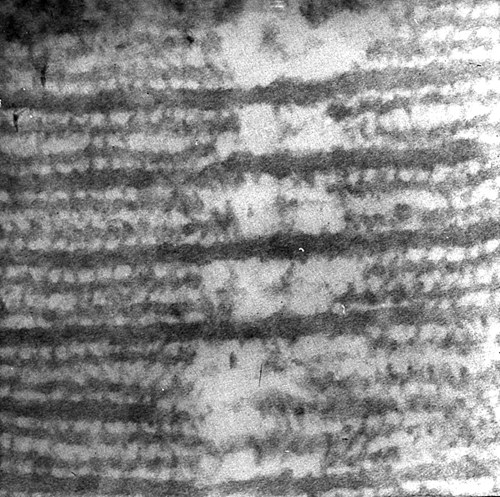}
}
\caption[cytoskeletonandcellmotility_fig2]
{\label{cytoskeletonandcellmotility_fig2}
Skeletal-muscle thick and thin filaments as seen by H. E. Huxley in 1957 \cite{HUXLEY:1957rt} (Reproductions from the original 1953 paper \cite{HANSON:1953fj} were poor). Left panel: Thin longitudinal sections of rabbit psoas muscle fibers, showing a single layer of a filament lattice, with individual thick and thin filaments as well as crossbridges between them. Right panel: Higher-magnification view of a thin longitudinal section. The relative dimensions were distorted due to axial compression during sectioning: crossbridges' axial spacing is $\simeq$ 40 nm and thick filaments' diameter is  $\simeq$ 12 nm. Source: reprinted from ref. \cite{Huxley:2004lr} with permission from Blackwell Publishing Ltd. (based on an original micrograph of 1957, see also ref. \cite{HUXLEY:1957rt}).}
\end{figure}

Cell movement was observed and reported for the first time as early as 1674, when Anthony van Leeuwenhoek brought a glass bead that served him as a primitive microscope close to a drop of water taken from a pool. His astonishment was immediate, as he later reported: ``.. the motion of these animalcules in the water was so swift and various, upwards, downwards and round about, that it was wonderful to see ...'' \cite{brayB}. The organisms he saw were probably ciliated \emph{protozoa} - unicellular non-photosynthetic eukaryotic organisms - a fraction of a millimeter in length, swimming by the agitated but coordinated motion of sometimes thousands of hairlike cilia on their surface (see Fig. \ref{cytoskeletonandcellmotility_fig1}).
Despite this very early observation, only relatively recent advances of the past few decades in microscopy, molecular biology and biochemistry have enabled the discovery of the basic underlying molecular mechanisms by which cells are able to feel their environment, exert forces and move in a directed way in search for nutrients or any other task they need to perform. The cytoskeleton, defined as the system of protein filaments that enable the cell to insure its structural integrity and morphology, exert forces and produce motion, was first observed by H. E. Huxley and J. Hanson in 1953, when they discovered the double array of filaments in cross-striated muscles using electron-microscopy techniques \cite{HUXLEY:1953yq,HANSON:1953fj,HUXLEY:1953kx}. In parallel with A. F. Huxley and R. Niedergerke, but independently, they published the next year the ``sliding-filament model", which explains muscle contraction via the relative sliding of two different types of filaments, originally called ``thick" and ``thin" filaments \cite{HUXLEY:1954fi,HUXLEY:1954vn} (see Fig. \ref{cytoskeletonandcellmotility_fig2}).
This, with the help of further genetic, biochemical and crystallographic studies, dated the beginning of a scientific understanding of the subcellular mechanisms that underly cell motility.

In addition to the characterization of the biochemical composition and organization of these subcellular structures, tremendous advances in the past two decades on both physical micro-manipulation and fluorescence-microscopy techniques have enabled the characterization of the processes involved with minute details. On the one hand, thanks to the help of micro-pipettes, atomic-force microscopes and optical tweezers, one can characterize the forces that cells are able to exert as well as their responses to applied stimuli. On the other hand, fluorescence-microscopy techniques provide information on the microscopic dynamics of single molecules \emph{in vivo}. Finally, combined with biochemistry and gene-expression control, as well as the micro-fabrication of bio-mimetic artificial systems in \emph{in vitro} assays, these techniques have enabled the study of simplified systems, where some specific aspects of the processes involved can be characterized separately.

In addition to these biochemical and behavioral characterizations, understanding the generic principles that underly cell motility needed an integrated approach to explain how this complex molecular machinery can self-organize and lead to a coherent, purposeful movement at the cellular level. Nothing better than a eukaryotic cell can indeed be categorized as a complex system, in that its behavior integrates the coordinated interplay of more than ten thousand different protein types, numbering together millions and representing 60\% of its dry mass \cite{cellB}. The cytoskeletal machinery is made of hundreds of different molecular players. For example, in 2003, about 160 proteins were known to bind to actin, one of the major biopolymers participating in cell structure and dynamical behavior \cite{Remedios:2003jt}. Knowledge about this biomolecular machinery is constantly evolving, and its undergoing complexity can be appreciated by consulting up-to-date information on available databases\footnote{See, for example, http://www.cytoskeletons.com/database.php.}. Therefore, understanding this complexity and describing how it is integrated at the cellular level was made by biophysical studies, both on experimental and theoretical grounds, which helped to identify the generic principles behind cell motility. At the molecular level first, the conversion of chemical energy stored in covalent bounds into mechanical work relies on out-of-equilibrium thermodynamic principles and asymmetrical properties - or polarity - of the structures involved, and happens in a highly-fluctuating environment of brownian particles \cite{Astumian:1997pd,reim02}. On larger length scales, the appearance of coordinated motion in large collections of proteins relies on collective phenomena, self-organization and dynamical symmetry breakings \cite{juli97b}. On yet larger length scales, swimming of microorganisms has attracted the attention of physicists for years \cite{tayl51,purc77}, and morphogenesis and pattern formations in cellular tissues rely on self-organization phenomena, as was envisioned for the first time by Turing in 1952 \cite{turi52,Kondo:2002bv}. Therefore, in addition to biophysical experimental techniques, variety of theoretical physics' disciplines spanning the theory of stochastic processes, statistical physics, out-of-equilibrium thermodynamics, hydrodynamics, nonlinear dynamics and pattern formation have contributed and still contribute to our understanding of cell motility.

The present article will mainly focus on the eukaryotic cytoskeleton and cell-motility mechanisms. Bacterial motility as well as the composition of the prokaryotic cytoskeleton will be only briefly mentioned. The article is organized as follows. In Section III, we will first present an overview of the diversity of cellular motility mechanisms, which might at first glance be categorized into two different types of behaviors, namely ``swimming'' and ``crawling''. Intracellular transport, mitosis - or cell division - as well as other extensions of cell motility that rely on the same essential machinery will be briefly sketched. In Section IV, we will introduce the molecular machinery that underlies cell motility - the cytoskeleton - as well as its interactions with the external environment of the cell and its main regulatory pathways. Sections \ref{Sse:OtherProteins} to \ref{Sse:Prokaryotes} are more detailed in their biochemical presentations; readers primarily interested in the theoretical modeling of cell motility might want to skip these sections in a first reading. We will then describe the motility mechanisms that rely essentially on polymerization-depolymerization dynamics of cytoskeleton filaments in Section V, and the ones that rely essentially on the activity of motor proteins in Section VI. Finally, Section VII will be devoted to the description of the integrated approaches that have been developed recently to try to understand the cooperative phenomena that underly self-organization of the cell cytoskeleton as a whole.

\section{\label{Se.CellMvt}The Diversity of Cell Motility}

\subsection{Swimming}

At the cellular level, viscous hydrodynamic forces are several orders of magnitude higher than inertial forces. Therefore, simple reciprocal motions cannot produce forward motion, and cellular swimming patterns need to be asymmetric in space and time for the cell to advance. This hydrodynamic problem faced by cells attempting to swim have been eloquently summarized by Purcell as ``life at low Reynolds number'' \cite{purc77}. To solve this problem, bacteria use the rotation of a short helical or corckscrew-shaped \emph{flagellum}, which is a relatively rigid structure made of a collection of hundreds of identical protein subunits called \emph{flagellins} \cite{Berg:2003uo}. The swimming of a single bacterium can be impressively rapid, as bacteria such as \emph{Escherichia coli} - the common intestinal bacterium - swim at speeds of 20 to 30 micrometers per second, for the cell itself is only about two-micrometer long and half a micrometer in diameter. The bacterium possesses multiple flagella that gather together during swimming, and can fly apart as the bacterium switches direction. Other bacteria such as \emph{Vibrio cholerae} - the causative agent of cholera - use a single flagellum located at one of their pole, but the propulsion mechanism relies always on the presence of a rotary molecular motor located in the cell membrane, and which is sensitive to modifications of the chemical environment of the cell. Under normal conditions, the bacterium changes direction in an intermittent chaotic way by reversing the rotational direction of its motors, a phenomenon known as \emph{tumbling}. When placed in a concentration gradient of nutrients however, the cell can adapt its tumbling frequency to swim towards nutrient-rich regions, a phenomenon known as \emph{chemotaxis} \cite{Manson:1998qr}.

Even though it shares the same name, the eukaryotic flagellum shares little structures and propulsion mechanisms with its bacterial counterpart. It is indeed at least ten times larger than a bacterial flagellum in both diameter and length, and instead of being a rigid passive structure animated by a remote motor, it bears its motor activity along its length. Propulsion occurs by the propagation of a bending wave along the flagellum as a result of the relative sliding of a group of about 10 long parallel filaments, which are engulfed in the cell's plasma membrane and are animated by hundreds of motor proteins in a coordinated manner. Eukaryotic cells also use another type of protrusions to swim, the \emph{cilia}, which are much like flagella in their internal structure, but which are shorter and work usually in numbers, covering sometimes the whole cell surface as in the case of paramecia or other ciliated protozoa (see Fig. \ref{cytoskeletonandcellmotility_fig1}). Their beating pattern is then coordinated at the cellular level, most often in a wave-type of manner known as the  \emph{metachronal wave}. Beating cilia are also found in animals, as for example in humans where ciliated cells play major roles in several organs like the brain, the retina, the respiratory tract, the Fallopian tube and the kidney \cite{Ibanez-Tallon:2003bd}.

Other strategies of swimming include the elegant movement used by \emph{Eutreptiella} - called \emph{metaboly} - which consists in gradually changing the contour of the cell surface to locally increase the drag exerted by the viscous fluid around and move the cell forward \cite{Fletcher:2004th}. Other organisms like most motile species of \emph{Chrysophytes} - a group of marine photosynthetic protozoa - possess a flagellum attached at their front instead of their back. The flagellum is covered with stiff hairs projecting from its side that allow the cell to move forward as a planar wave propagates from the base to the tip of the flagellum \cite{brayB}. Finally, one should mention yet another type of motility - namely \emph{walking} - in which cells use also cilia and flagella animated in a coordinated manner to enable the cell to literally ``walk'' over surfaces. Walking motility relies on the same essential biochemical structures as the ones employed in swimming with collections of cilia.

\subsection{Crawling}

Cell crawling is the common mechanism employed by most eukaryotic animal cells as they move through animal tissues, constituted of other cells or filaments of the extracellular matrix \cite{cellB,brayB}. In contrast to swimming cells, crawling cells in general do not employ conspicuous motile organelles that are external to the cell, and which can be studied in isolation. In general however, they either move by means of wormlike cycles of extensions and contractions of their cell body or of some specific protrusions, or slide without visible means of protrusion, a process also referred to as \emph{gliding}. Most of crawling mechanisms rely on the protrusion of specific dynamic extensions at the leading edge of the cell, but gliding seems to rely sometimes on different mechanisms \cite{Heintzelman:2003xe,Fletcher:2004th}. Even though gliding mechanisms are widespread in bacteria, algae and parasitic protozoa, we still do not know for sure the molecular machinery as well as the essential mechanisms that underly these different phenomena \cite{brayB,Heintzelman:2003xe}. 

\begin{figure}[h]
\scalebox{0.12}{
\includegraphics{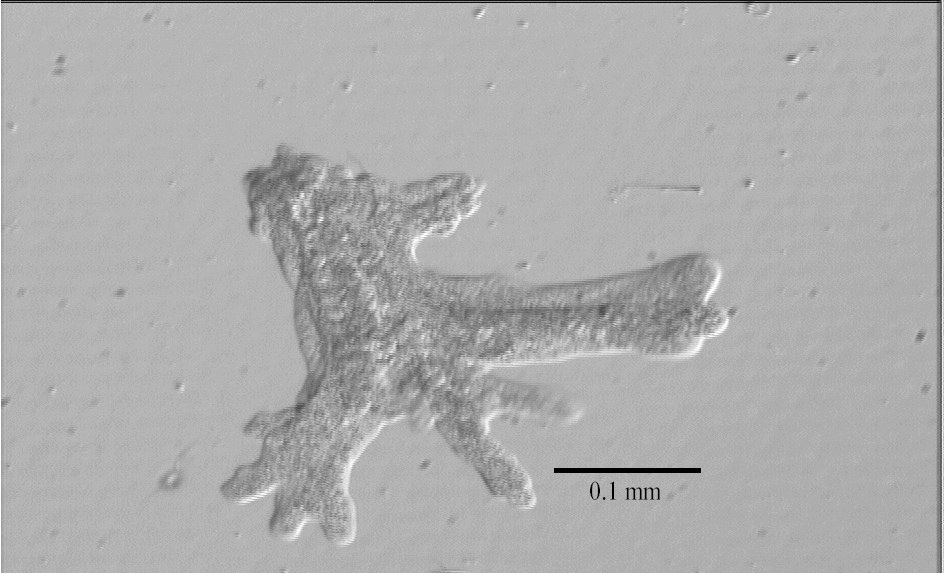}
}
\scalebox{0.12}{
\includegraphics{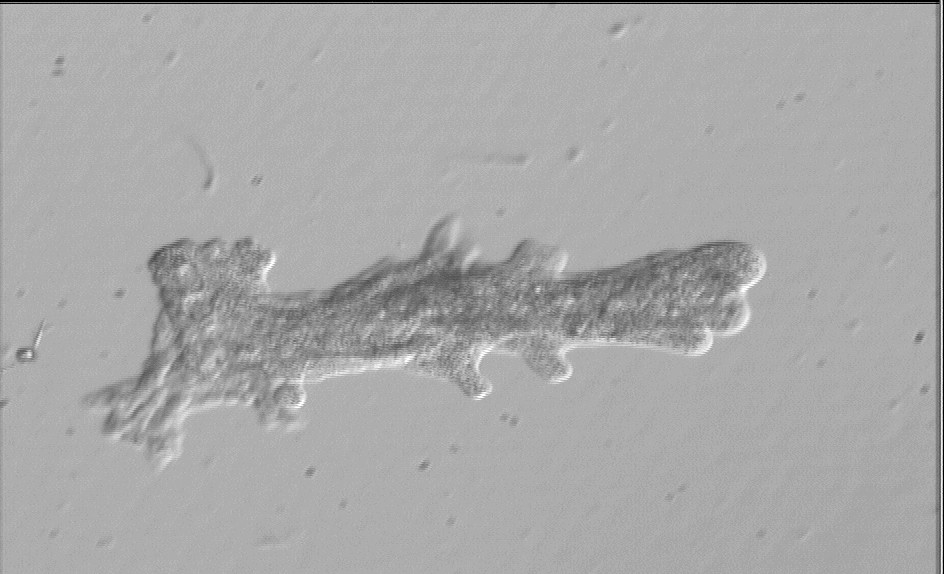}
}
\caption[cytoskeletonandcellmotility_fig3]
{\label{cytoskeletonandcellmotility_fig3}
Two pictures of \emph{amoeba proteus} displaying different shapes of its pseudopodia. Note the dramatic change in cell shape during locomotion. Source: courtesy of Sutherland Maciver.}
\end{figure}

\begin{figure}[h]
\scalebox{0.17}{
\includegraphics{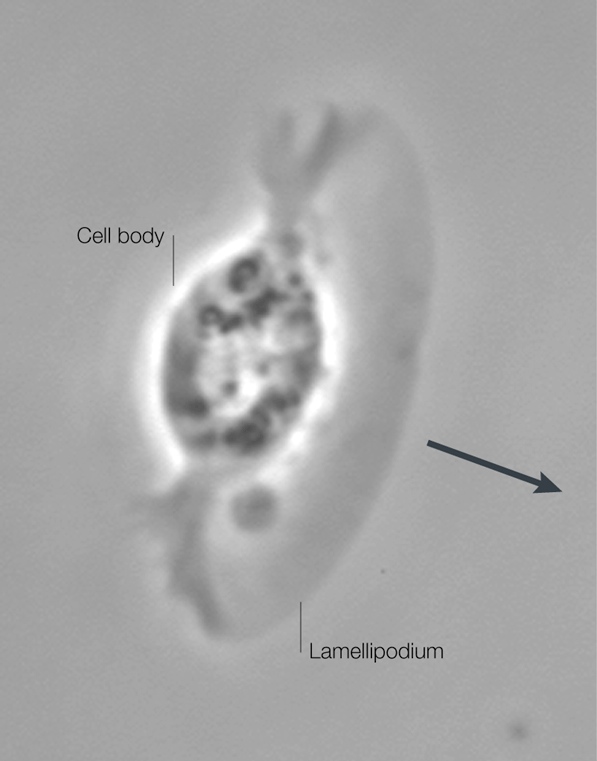}
}
\scalebox{0.4}{
\includegraphics{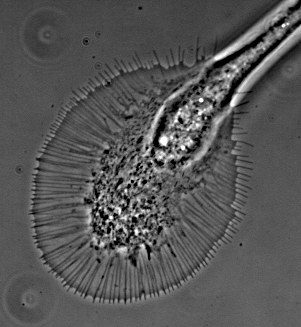}
}
\caption[cytoskeletonandcellmotility_fig4]
{\label{cytoskeletonandcellmotility_fig4}
Two examples of lamellipodia. Left panel: A living fish keratocyte extends its leading lamellipodium during crawling. This is a phase-contrast micrograph, a single frame from a video sequence\footnote{The whole movie can be seen on the following webpage: http://cmgm.stanford.edu/theriot/movies.htm.}. The lamellipodium and the cell body are labelled. The large arrow indicates the direction of motion. Source: reprinted from ref. \cite{Cameron:2000rr} with permission from Nature Publishing Group. Right panel: Snail neuronal growth cone by means of which the nerve fiber elongates at its tip. Clearly visible are the radially-aligned bundle structures that project into filopodia at the leading edge of the lamellipodium. Source: courtesy of Feng-quan Zhou; reprinted from ref. \cite{Zhou:2001lr} with permission from Rockefeller University Press.}
\end{figure}

The best characterized crawling mechanism is the so-called \emph{amoeboid motility}, referring to the locomotion of all eukaryotic cells that move by means of protrusion of their leading edge. Originally, the term was referring uniquely to the crawling mechanism of \emph{Amoeba proteus}, a particular species of \emph{amoebae}, whose protrusions are stubby three-dimensional projections called \emph{pseudopodia}\footnote{Note that some zoologists also use the term ``pseudopodia'' or ``pseudopods'' rather generally to refer to a variety of cell-surface protrusions. These include the different types of protrusions described here as playing a role in amoeboid motility, but also the long extended processes that some cell types use only as feeding apparatus, like \emph{axopodia}.} (see Fig. \ref{cytoskeletonandcellmotility_fig3})\footnote{A short video of a locomoting \emph{Amoeba proteus} can be seen on the following website: http://www.bms.ed.ac.uk/research/others/smaciver/A.prot.Loc.mov.}.
But other types of protrusions exist, that are classified with respect to their shape and dimensional organization. Two-dimensional protrusions are the flat veil-shaped projections called \emph{lamellipodia}, as they occur in fibroblasts' or fish epidermal keratocytes' motility for wound healing\footnote{Fibroblasts are the cells that synthesize and maintain the extracellular matrix in most animal connective tissues. They provide a structural framework (stroma) for many tissues, and play a crucial role in wound healing. Keratocytes are epithelial cells that have been characterized in the epidermis of fish and frogs, and that have been named so because of their abundant keratin filaments. They are specialized in wound healing, and are one of the most spectacular example of fast and persistent locomotion in cells, with velocities up to 30 $\mu$m/min \cite{cellB,Lee:1993lr}.}. One-dimensional projections are the long thin projections called either \emph{filopodia} or \emph{microspikes}, and which occur for example in neuronal-growth-cone progressions\footnote{Growth cones are structures that are found at the tip of axons and dendrites, by means of which neuron cells extend.}. Filopodia usually protrude as small extensions of a lamellipodium, and are used by the cell to extend its lamellipodium in a given direction \cite{Small:2002sw,Kaverina:2002bs} (see Fig. \ref{cytoskeletonandcellmotility_fig4}).
To these must be added the spherical membrane protrusions called \emph{blebs}, which occur as a result of cortical contractility, and which have been proposed recently to participate in the initiation of lamellipodium formation and the elaboration of cell polarity for directed motion \cite{Paluch:2006fj}, as well as in the amoeboid motility itself for example in \emph{Dictyostelium}, a model species of \emph{amoebae} \cite{Yoshida:2006cl}. Finally, one should mention  a motility mechanism that can be classified as \emph{rolling}, in which some organisms such as \emph{helizoa} use coordinated shortening and lengthening of long radiating needlike extensions called \emph{axopodia} to roll over surfaces. Axopodia happen also to be sticky extensions that are most often used for catching preys in numerous protozoa.

Depending on authors, the process of amoeboid motility can be decomposed into three to five steps that occur simultaneously. First the cell makes a protrusion, where the membrane is pushed forward by means of the polymerization of cytoskeletal filaments. Then the protrusion adheres to the substrate via the formation of anchoring points, and subsequent contraction of the cell cytoskeleton drags the cell body forward. Finally at the rear end, the cell de-adheres and retracts \cite{Lauffenburger:1996cr,Mitchison:1996ul}. Of these five steps, the two last ones - namely de-adhesion and retraction - involve similar structures and mechanisms as the formation of the anchoring points and cell body drag, which led originally Abercrombie to describe his observation as a three-step cycle \cite{aber80}. The speed of amoeboid motility can range from less than a micrometer per hour to more than one micrometer per second, depending on cell type and stimulation\footnote{Cell-motility videos can be seen at http://cellix.imba.oeaw.ac.at.}.

\subsection{\label{Sse.ExtCellMot}Extensions of cell motility}

\begin{figure}[h]
\scalebox{0.18}{
\includegraphics{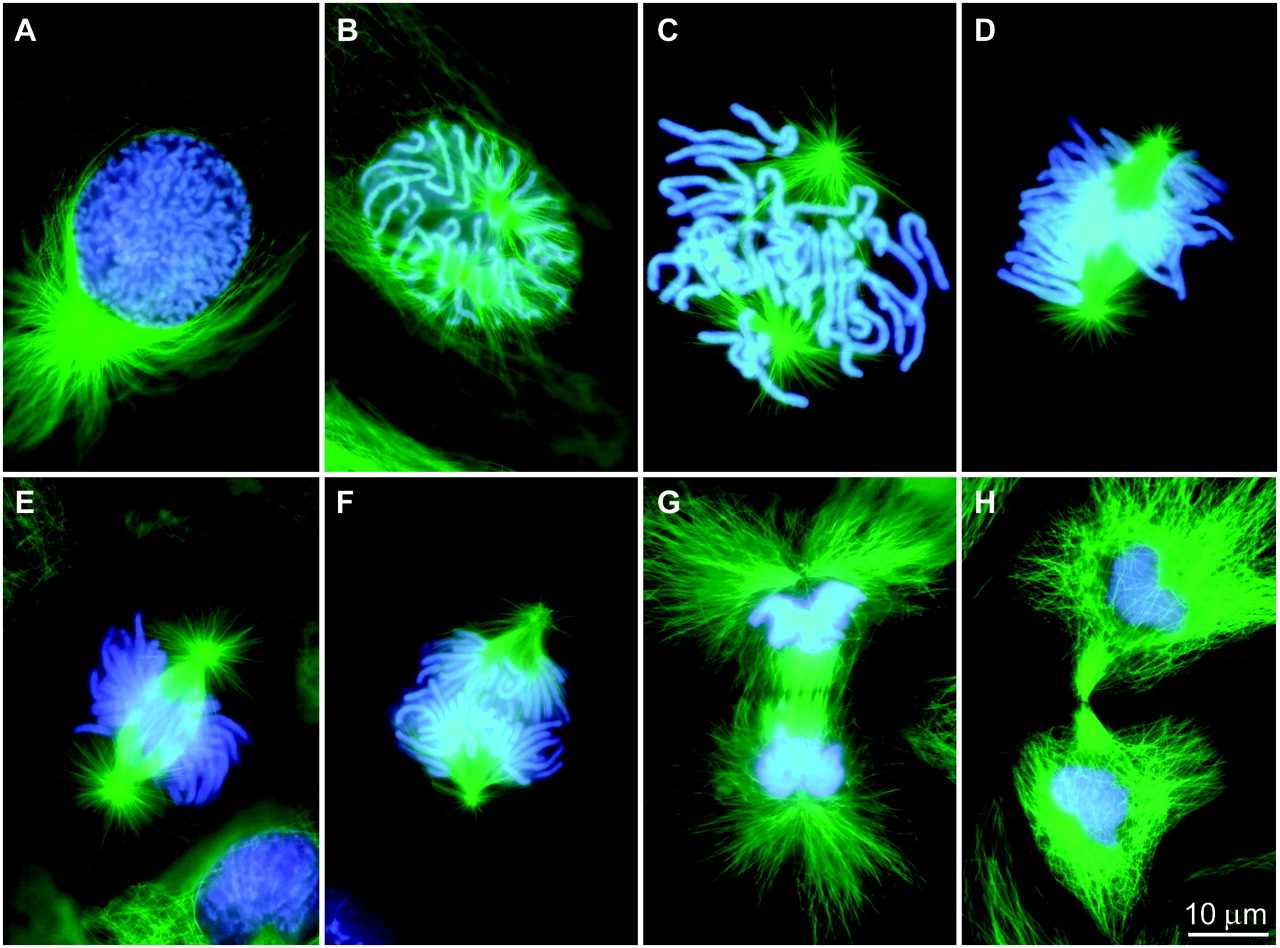}
}
\caption[cytoskeletonandcellmotility_fig5]
{\label{cytoskeletonandcellmotility_fig5}
(A to H) Fluorescence micrographs of mitosis in fixed newt lung cells stained with antibodies to reveal the microtubules (MT, green)\footnote{see Section \ref{Sse.Biopolymers}.}, and with a dye (Hoechst 33342) to reveal the chromosomes (blue). The spindle forms as the separating astral MT arrays, associated with each centrosome (A to C), interact with the chromosomes. Once the chromosomes are segregated into daughter nuclei (F and G), new MT-based structures known as stem-bodies form between the new nuclei (G). These play a role in cytokinesis (H), the actual cleavage of the two daughter cells. Source: reprinted from ref. \cite{Rieder:2003fk} with permission from The American Association for the Advancement of Science.}
\end{figure}

In addition to moving the whole cell body, the machinery that is responsible for cellular movement can be employed for quite different tasks, which are as essential to the cell survival and reproduction as its motility \emph{per se}. As we have earlier stated, even in the case of macroscopically non-aminated live forms, the constitutive cells need constantly to displace their internal organelles for their metabolism to be maintained \cite{Rogers:2000fj}. When looked under the light microscope, mitochondria, vesicles, lysosomes and ingested particles display a rapid and sporadic movement that is interspaced with relatively long periods of quiescence. Velocities are typically of the order of micrometers per second, as the fastest known organelle transport is performed in the green algae \emph{Chara}, whose chloroplasts are transported at velocities that can achieve 60 micrometers per second \cite{Chaen:1995hc}. Of all cell types, the need for organelle transport is best illustrated by the mammalian motor neurons whose longest extensions - the axons - even though typically only a few micrometers in diameter, can reach lengths up to one meter. Characteristic times that would be required for a mitochondrion to naturally diffuse that distance in such a geometry range from 10 to 100 years. Instead, membrane vesicles and organelles are actively transported in both directions at speeds of about one to five micrometers per second, which allows the whole journey to be made in just a few days \cite{brayB}. Finally, the probably most-spectacular event of intracellular transport occurs during the essential process of eukaryotic mitosis, by which duplicated chromosomes are segregated from the mother cell and delivered to each of the nascent daughter cells. For this process to occur, major structural reorganizations of the whole-cell cytoskeleton are needed, during which a large and complex cellular structure - the \emph{mitotic spindle} - assembles and drives the chromosomes apart in a coordinated manner \cite{Mitchison:2001qe} (see Fig. \ref{cytoskeletonandcellmotility_fig5}).

Cell motility can also occur by means of molecular machineries located outside the cell that needs to move. This is the case in particular for mammalian pathogene bacteria such as \emph{Listeria monocytogenes} and \emph{Shigella} and \emph{Rickettsia} species, but also for some viruses like \emph{vaccinia virus}. These organisms propel themselves within and across the cells they invade by utilizing the cytoskeleton of their hosts \cite{Tilney:1989kl,Cossart:2001zm,Goldberg:2001ec,Gruenheid:2003zv}. Among these organisms, \emph{Listeria} in particular has become a model organism for studying actin-based motility, a simplified version of the whole amoeboid motility, and which can be seen as a representation of just the first step of this complex process in the original Abercrombie description \cite{Cameron:2000rr,Plastino:2005yq,Mogilner:2006dz}. Other particular systems use different specific structures from purely cell-cytoskeleton-based motility. Among these, vertebrate skeletal muscles contrast with standard cellular motility, in that they are structured in enormous multinucleated cells that evolved specifically to generate extremely rapid, repetitive and forceful movements. The cytoplasm of these giant cells is crammed full of a highly-organized, almost-crytsalline array of cytoskeletal filaments, whose only function is to produce contractile forces \cite{brayB,Huxley:1996si}. Another essential event of cell division, namely \emph{cytokinesis} - the actual cleavage of the two daughter cells - also involves the contraction of relatively-sliding cytokeletal filaments, this time under the form of a dividing ring \cite{Glotzer:2001oq}.

Finally, other types of cells use mechanisms that do not rely on their cytoskeleton for their motility. Of the most spectaculars is the motility based on the stored mechanochemical energy in some supra-molecular springs, which can then contract at velocities as high as eight centimeters per second \cite{Mahadevan:2000cr}. Yet another mechanism relies on some stored purely-elastic energy that allows some insect-eating plants to catch their preys. This is the case for example of the Venus flytrap \emph{Dionaea muscipula}, whose leafs can close in about 100 ms, one of the fastest movements in all plant kingdom. To achieve such a performance, the plant relies on a snap-buckling instability, whose onset is actively controlled by the plant after the arrival of a fly has triggered some biochemical response via the disturbance of mechano-sensitive hairs located inside the trap \cite{Forterre:2005kh}.

\section{\label{Se.Cytoskeleton}The Cell Cytoskeleton}

The eukaryotic \emph{cytoskeleton} is defined as the system of protein filaments that enable the cell to insure its structural integrity and rigidity, regulate its shape and morphology, exert forces and produce motion. As a framework that insures structural integrity, the cytoskeleton is mainly constituted of a cohesive meshwork of protein filaments that extend throughout the cytoplasm of the cell. But being the essential structure that produces movement at the cellular level, and thereby needing to be highly adaptable to extracellular stimuli or rapid environmental changes, the cytoskeleton has evolved into a highly-dynamic structure. In fact, cytoskeletal filaments constantly grow and shrink, associate and dissociate via multiple linkages, organize on large scales into a dynamic network, and serve as an intricated set of tracks to motor proteins that transport cargos from one part of the cell to the other, or slide filaments with respect to one another to produce contractile forces. This section is devoted to the biochemical description of this very-complex structure. In addition, its interaction with the cell's external world and its regulatory pathways will be briefly presented, as well the prokaryotic cytoskeleton which, even though biochemically different, appears more and more to resemble its eukaryotic counterpart on a functional point of view.

\subsection{\label{Sse.Biopolymers}Biopolymers}

\begin{figure}[h]
\scalebox{0.4}{
\includegraphics{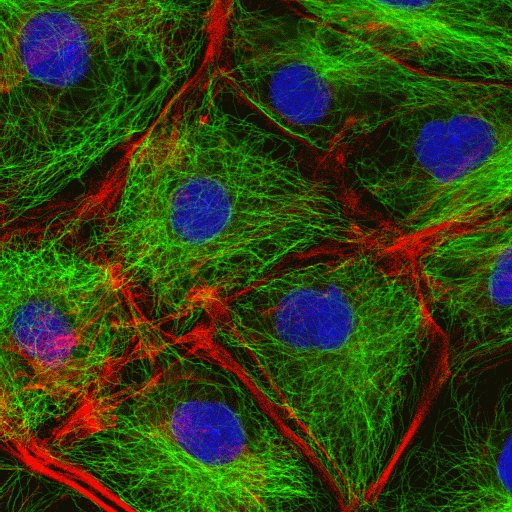}
}
\caption[cytoskeletonandcellmotility_fig6]
{\label{cytoskeletonandcellmotility_fig6}
Animal cells as seen in the fluorescence microscope after fixation and labelling with specific probes. Actin filaments are stained in red, microtubules in green, and the nuclei in blue. Source: courtesy of Mark Shipman, James Blyth and Louise Cramer,
MRC-Laboratory Molecular Cell Biology and Cell Biology Unit, UCL, London UK (unpublished).}
\end{figure}

How can a eukaryotic cell, with a diameter of 10 microns or more, be spatially organized by cytoskeletal proteins that are typically 2000 times smaller in linear dimensions ? The answer lies in \emph{polymerization}, this ability of the elementary protein subunits (called \emph{monomers}) to assemble via physical interactions into extended linear structures that are made of a large number of them, typically thousands (called thereby \emph{polymers}, or here more precisely \emph{biopolymers}). There are three types of biopolymers in a given eukaryotic cell, namely \emph{actin filaments}, \emph{microtubules} and \emph{intermediate filaments} (see Fig. \ref{cytoskeletonandcellmotility_fig6}).
Although they are classified according to their respective thickness, more interesting for cellular structures and functions are their rigidity, which at thermodynamic equilibrium is characterized by their persistence length $L_p$\footnote{The persistence length $L_p$ is defined as follows: consider a thin flexible rod of fixed length $L$, submitted to thermal forces. Its shape is completely specified by the tangent angle $\theta(s)$ in three dimensions along the arc length of the rod \cite{LandauLifschitzElasticity}. The persistence length $L_p$ is defined as the characteristic arc length above which thermal fluctuations of the angle $\theta(s)$ become uncorrelated. Specifically, $\langle \cos{[\Delta\theta(s)]}\rangle = \exp{(-s/L_p)}$, where $\Delta\theta(s)$ is the three-dimensional angle change over the arc length $s$. $L_p$ is related to the rod's material Young modulus $E$ and its geometrical moment of inertia $I$ by $L_p=EI/k_B T$, where $k_B T$ represents thermal energy \cite{Gittes:1993ly}.} \cite{Gittes:1993ly}.

\begin{figure}[h]
\scalebox{0.65}{
\includegraphics{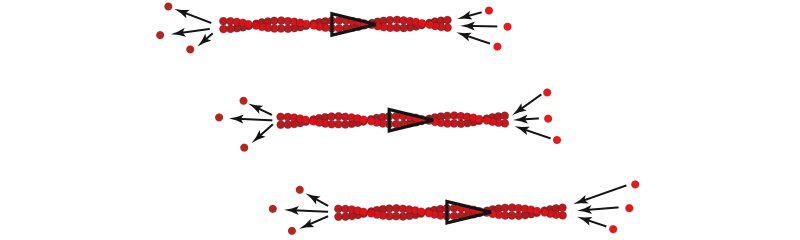}
}
\caption[cytoskeletonandcellmotility_fig7]
{\label{cytoskeletonandcellmotility_fig7}
Schematic representation of a treadmilling actin filament. The arrows indicate the polarity of the filaments. Monomers are added to the plus end and removed from the minus end at the same rate, such that while the filament's length remains constant, its center of mass is advancing. Top to bottom shows three subsequent times. Source: courtesy of Karsten Kruse.}
\end{figure}

Actin filaments - or F-actin - have a persistence length that is usually accepted to be of the order of 15 to 17 $\mu$m \cite{Ott:1993ys,Gittes:1993ly}, even though it has been reported that actin rigidity should depend on the way it is decorated, ranging from $9 \pm 0.5$ $\mu$m for bare F-actin to $20 \pm 1$ $\mu$m for tropomyosin-bound actin filaments in skeletal-muscle structures \cite{Isambert:1995fp}. Actin filaments are two-stranded helical polymers, 5 to 9 nm in diameter, and are built from dimer pairs of globular-actin monomers - or G-actin - that are polar in nature \cite{cellB,brayB}. The two halves of an actin monomer are separated by a cleft that can bind adenosine triphosphate (ATP) or its hydrolyzed form adenosine diphosphate (ADP)\footnote{The hydrolysis reaction of ATP (ATP $\rightleftharpoons$ ADP + $\rm{P}_{\rm{i}}$, where $\rm{P}_{\rm{i}}$ designates inorganic phosphate) breaks a high-energy chemical bond - here a phosphoanhydride bond - to drive many chemical reactions in the cell.}. This is responsible for the existence of two distinct ends to the whole filament, namely a fast growing end - called ``plus end'' or ``barbed end'' - where mostly ATP-bound monomers are located, and a slow growing end - known as ``minus end'' or ``pointed end'' - that is rich in ADP-bound monomers. The minus end has a critical actin-monomer concentration that is roughly six times as high as that of the plus end. At steady state, and with the help of monomeric diffusion, this drives the phenomenon of \emph{treadmilling}, a dynamic evolution of the actin filament where actin monomers are added to the plus end, and removed from the minus end at the same rate. During this process, the total length of the treadmilling filament is kept constant, while its center of mass is displaced at a constant velocity, even though each individual polymerized monomer do not move on average\footnote{Animated movies of this process can be seen at http://www.uni-leipzig.de/~pwm/kas/actin/actin.html or http://cellix.imba.oeaw.ac.at/actin-polymerisation-drives-protrusion.} (see Fig. \ref{cytoskeletonandcellmotility_fig7}).
For pure actin at physiological concentrations, this process is rather slow and occurs at velocities of the order of a few micrometers per hour. But as we shall see in the following, specialized actin-binding proteins allow the cell to increase this speed substantially, which makes actin forces exerted via polymerization-depolymerization mechanisms one of the key players in cellular motility.

Actin is the most abundant protein in a eukaryotic cell (several grams per liter), and has been highly conserved throughout evolution. It organizes into a variety of structures, namely linear bundles, two-dimensional networks or three-dimensional gels, and is mainly concentrated in a layer located just beneath the plasma membrane and called the \emph{actin cortex}. Of primary importance for cell motility are the two-dimensional highly cross-linked networks that actin forms in lamellipodia, and the linear bundles that are found in filopodia and which protrude from the lamellipodia in a directed way. There, as we shall see in Section V, actin polymerization plays a crucial role in driving cell motility. Finally, one should mention the cortical rings that contract during the process of cytokinesis to cleave the two daughter cells, as well as the formation of \emph{stress fibers}, which are force-producing structures that are attached to anchoring points, and which enable the cell to exert traction forces on the substrate on which the cell is crawling\footnote{Illustrations of these structures can be found at http://cellix.imba.oeaw.ac.at. See also \cite{Small:2002sw,Kaverina:2002bs}.} (see Section \ref{Sse:CellAnchoring}).

Microtubules are the stiffest of all polymers, with persistence lengths ranging from 100 $\mu$m up to 6 mm \cite{Pampaloni:2006ek}. They are hollow cylinders with an outer diameter of $\simeq$ 25 nm and are made of tubulin subunits arranged in 13 adjacent protofilaments. Tubulin is a heterodimer formed of $\alpha$- and $\beta$-subunits, which can bind either guanosine triphosphate (GTP) or guanosine diphosphate (GDP)\footnote{Similarly to ATP, GTP is a stored source of energy for the cell that is consumed via a hydrolysis reaction, here GTP $\rightleftharpoons$ GDP + $\rm{P}_{\rm{i}}$.}. Microtubules share some important properties with actin filaments, in that they are polar, treadmill, and can exert forces \cite{Dogterom:2005vl}. They typically organize radially from a single microtubule-organizing center called the \emph{centrosome}, and connect to the actin cortex with their plus ends towards the cell edge. In addition to giving the cell its structural rigidity and shape, they actively participate in regulating the actin cortex dynamics, focal-adhesion assembly and disassembly, and in some cell types participate in determining the cell polarity and its subsequent migrating direction (see Section \ref{Sse:CellAnchoring}).

Intermediate filaments are the most flexible polymers of the cell cytoskeleton, with persistence lengths of the order of 0.3 to 1.0 $\mu$m. They range in diameter from 7 to 12 nm, in-between that of actin and microtubules. There are different classes of intermediate filaments such as \emph{vimentin}, \emph{desmin}, \emph{keratin}, \emph{lamin} and \emph{neurofilaments}, and they constitute together a large and heterogeneous family, of which different cells possess different members. Unlike actin filaments and microtubules, they are not polar, do not treadmill, and are therefore thought to contribute essentially to the structural and elastic properties of the cell, but little to its dynamics and motility. One particular example of intermediate-filament structure is the \emph{nuclear lamina}, located just beneath the inner nuclear membrane, and that is responsible for its structural integrity.

\subsection{\label{Sse.Motors}Molecular motors}

\emph{Molecular motors} constitute the subset of proteins and macromolecular complexes that convert a given source of energy into mechanical work \cite{cellB,howaB}. The energy they need is generally stored into either of two forms by the cell: high-energy chemical bonds, such as the phosphoanhydride bonds found in ATP and GTP, and asymmetric ion gradients across membranes. Known molecular motors can be classified into roughly five categories, namely (1) rotary motors, (2) linear-stepper motors, (3) assembly-disassembly motors, (4) extrusion nozzles, and (5) prestressed springs. A nice table of the major different cell-movements' categories with the different cellular structures and molecular motors they rely on, can be found in ref. \cite{Fletcher:2004th}.

All known biological rotary motors use ion-gradient-based sources of energy, and most of them use electrochemical forces based on hydrogen-ion (or proton) gradients, also known as \emph{proton-motive forces}. This is the case for example for the propulsion motor of bacteria that is responsible for their flagella to rotate \cite{Manson:1998qr,Berg:2003uo}, as well as for the surprising rotary motor F0F1-ATPase that is responsible for ATP synthase in mitochondria and bacteria \cite{Yoshida:2001kl}. This rotary machine usually converts the electrochemical energy stored in proton-concentration gradients, first into mechanical motion, and then back into chemical energy under the form of ATP. But the motor is also reversible, in that it can harness the chemical energy of ATP to produce or maintain the transmembrane electrochemical gradient of proton concentration. This reversibility is best seen in bacteria, when they switch from aerobic to anaerobic conditions \cite{cellB}.

\begin{figure}[h]
\scalebox{0.38}{
\includegraphics{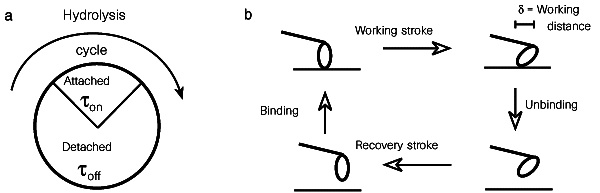}
}
\caption[cytoskeletonandcellmotility_fig8]
{\label{cytoskeletonandcellmotility_fig8}
(a) ATP-hydrolysis cycle, with the respective durations $\tau_{\rm{on}}$ and $\tau_{\rm{off}}$ of the attached and detached states of the motor. These durations define the duty ratio $r$ as $r=\tau_{\rm{on}}/(\tau_{\rm{on}}+\tau_{\rm{off}})$. (b) During the attached phase, the head of the motor makes a working stroke of working distance $\delta$. The motor then unbinds from the filament, and makes a recovery stroke during the detached phase. By recovering its initial conformation while detached, the motor avoids stepping backwards and so progresses by a distance equal to the working stroke during each cycle. Source: reprinted from ref. \cite{Howard:1997bh} with permission from Nature Publishing Group.}
\end{figure}

Most of the motors used in amoeboid motility are linear-stepper motors \cite{howaB,Spudich:1994ud,Schliwa:2003yb}. We shall therefore focus on this class of motor proteins in the remaining of the present article. They walk on the linear tracks formed by the polymerized cytoskeletal filaments, and can be classified into two different categories, namely \emph{processive} and \emph{non-processive} molecular motors, sometimes designated as ``porters'' and ``rowers'' \cite{Leibler:1993ph}. The processivity is linked to the \emph{duty ratio}, the proportion of time that the molecule spends attached to the filament as compared to the whole motor cycle, namely one ATP-hydrolysis cycle \cite{Howard:1997bh} (see Fig. \ref{cytoskeletonandcellmotility_fig8}).
Typically, porters are individual walkers that carry cargos across the cell, and therefore most often participate in intra-cellular traffic. Rowers however work in numbers, and are usually involved in generating contractile forces, like it is the case in skeletal-muscle fibers, stress fibers or contractile rings that form during cytokinesis \cite{HUXLEY:1954fi,HUXLEY:1954vn,Huxley:1996si,Glotzer:2001oq}. Structurally, all these motor proteins can be divided into a motor domain, called the \emph{head}, and a \emph{tail} or \emph{base}. The head is the site of conformational change of the protein during ATP-hydrolysis, and with which the motor attaches to the filament. The tail connects the motor to its cargo or to other motors. Processive motors are (homo-)dimers, such that as one head is attached to the filament, the other can move to a new binding site. In that case, the two tails of the associated monomers wind up together to hold to each other. Non-processive motors can also be found in dimeric forms, one of the two heads being then just unused.

\subsection{\label{Se.MotorFamilies}Motor families}

\begin{figure}[h]
\scalebox{0.8}{
\includegraphics{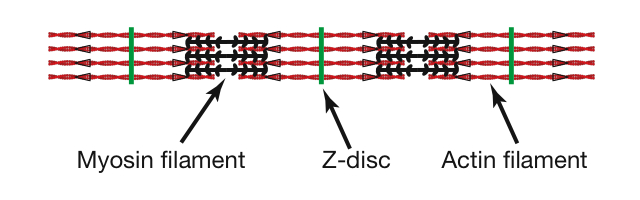}
}
\caption[cytoskeletonandcellmotility_fig9]
{\label{cytoskeletonandcellmotility_fig9}
Schematic representation of muscle myofibrils, the basic contractile fibers of skeletal muscles. Actin and myosin filaments are periodically arranged in a polarity-alternated fashion. Between two ``Z discs'' is found the elementary structure that is periodically repeated, the \emph{sarcomere}, and where relative sliding of actin and myosin filaments leads to contraction. Source: courtesy of Karsten Kruse.}
\end{figure}

Eukaryotic cytoskeletal motor proteins are divided into three superfamilies, namely \emph{myosins}, \emph{kinesins} and \emph{dyneins}. The motor proteins known longest belong to the myosin superfamily \cite{Berg:2001eu}, because of their high concentration in skeletal muscles. All myosin motors walk on actin filaments through a general four-step process: binding, power-stroke, unbinding, and recovery-stroke\footnote{Animated movies of myosin skeletal fibers' detailed motion can be seen at http://www.scripps.edu/cb/milligan/research/movies/myosin.mov, or http://valelab.ucsf.edu.} \cite{Howard:1997bh} (see Fig. \ref{cytoskeletonandcellmotility_fig8}). Today, they are classified into 18 different classes, with possibly dozens of different members in each class, even in a single organism. The skeletal-muscle myosins belong to the Myosin II family; they have long tails that form dimeric $\alpha$-helices and associate into the so-called ``thick filaments'' originally observed by H. E. Huxley and J. Hanson, while the ``thin filaments'' are F-actin polymers \cite{HUXLEY:1953yq,HANSON:1953fj,HUXLEY:1953kx} (see Fig. \ref{cytoskeletonandcellmotility_fig2} and \ref{cytoskeletonandcellmotility_fig9}).
Most myosin molecules are plus-ended directed (to the exception of Myosin VI), and non-processive (to the exception of Myosin V, which is involve in vesicular transport). Their very diverse mechanical features, in terms of step sizes, duty ratios and stepping speeds, are very fine-tuned to their functions\footnote{Up-to-date information about myosin motors can be found at http://www.proweb.org.} \cite{Howard:1997bh}.

Kinesin proteins share very similar structural features with myosins in their head domain and are therefore thought to have branched from a common ancestor with myosins, but diverge in their tail structures \cite{Vale:2000nx}. They walk on microtubules instead of actin filaments, are processive, and are involved mainly in intracellular transport like the transport of organelles along nerve axons\footnote{Animated movie of kinesin's detailed motion can be seen at http://www.scripps.edu/cb/milligan/research/movies/kinesin.mov or http://valelab.ucsf.edu.}. The kinesin superfamily has been divided into 14 families, and a number of ``orphans'' that are so far ungrouped \cite{Miki:2001lq}. Most kinesin motors are plus-ended directed, like the conventional kinesin I that founded the family \cite{Vale:1985fu}. Members of the Kinesin-13 family are unconventional, in that they can processively induce microtubule depolymerization, a process that is essential to chromosome segregation during mitosis\footnote{Up-to-date information about kinesin motors can be found at http://www.proweb.org.} \cite{Hunter:2003wt} (see next Section).

Dynein proteins are less well-characterized. It is also unknown whether they share a common ancestor with myosins and kinesins, or whether they are the result of convergent evolution. Two major groups of dyneins exist: axonemal dyneins, which drive the bending of eukaryotic cilia and flagella by inducing the relative sliding of microtubules \cite{Porter:1996mw}, and cytoplasmic dyneins, which are involved in organelle and vesicular transport, as well as cell division \cite{Karki:1999cz}. Most dyneins are minus-ended directed, and interestingly, some dyneins can be non-processive at high, but processive at low ATP concentrations.

\subsection{\label{Sse:OtherProteins}Other cytoskeleton-associated proteins}

\begin{figure}[h]
\scalebox{0.4}{
\includegraphics{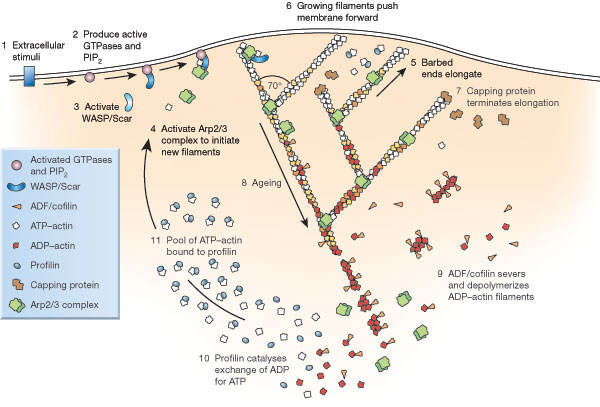}
}
\caption[cytoskeletonandcellmotility_fig10]
{\label{cytoskeletonandcellmotility_fig10}
Dynamical organization of the actin network at the leading edge of a protruding lamellipodium. (1) External cues activate signalling pathways that lead to GTPases and $\textrm{PIP}_{\rm{2}}$ activation (2). These then activate proteins of the WASP family (3), which in turn activate Arp2/3 complexes that initiate new filaments as branching from existing ones (4). Each new filament grows rapidly (5), fed by a high concentration of profilin-bound actin stored in the cytoplasm, and this pushes the plasma membrane forward (6). Capping proteins bind to the growing ends, terminating elongation (7). Actin-depolymerizing factor (ADF)/cofilin sever and depolymerize the ADP filaments, mainly in the ``older regions'' of the filaments (8, 9). Profilin re-enters the cycle at this point, promoting dissociation of ADP and binding of ATP to dissociated subunits (10). ATPÐactin binds to profilin, refilling the pool of subunits available for assembly (11). Source: reprinted from ref. \cite{Pollard:2003lr} with permission from Nature Publishing Group (image based on an original figure of ref. \cite{Pollard:2000wd}).}
\end{figure}

The coordination of the numerous different processes that happen during amoeboid motility rely on a tight regulation of the activity of the cell cytoskeleton, as well as its anchoring to the substrate. In particular, as we shall see below, the protrusion of the leading edge of the cell - the first step of amoeboid motility - relies on the formation of a highly-cross-linked and dynamic network of actin filaments. Its formation and dynamical regulation are carried out with the help of numerous accessory proteins \cite{Pollard:2000wd,Small:2002sw}. Following Pollard's presentation \cite{Pollard:2003fh,Pollard:2003lr} (see Fig. \ref{cytoskeletonandcellmotility_fig10}), we can focus on the main proteins that are involved in the formation, structure and dynamics of the actin network.
Nucleation of the network starts after biochemical signals have been integrated via G-protein-linked membrane receptors, namely small GTPases and $\textrm{PIP}_2$ pathways \cite{Machesky:1999fv} (see Section \ref{Sse:CellAnchoring}). Then members of the Wiscott Aldrich syndrome protein (WASP) family that are anchored to the cell's plasma membrane (like Scar \cite{Machesky:1999fy}), activate Arp2/3 (for actin-related proteins 2 and 3) complexes that are responsible for the nucleation and maintenance of branching points in the network\footnote{In addition to Arp2 and Arp3, which are members of the Actin related proteins (Arp) family in that they have sequences and structures that are similar to actin, the Arp2/3 complex contains five other smaller proteins.}. Then, in order to promote growth of the actin gel, recycling of G-actin monomers as well as the creation of new F-actin plus ends are stimulated by mainly two types of proteins: (1) Actin-binding proteins - such as profilin - that bind to actin monomers, catalyze the exchange of ADP for ATP, and inhibits ATP hydrolysis, a process that is antagonized by monomer-sequestering proteins - like thymosine$\beta$-4 - that stabilize ADP-bound G-actin. (2) Actin-depolymerizing factors (ADF) - such as cofilin (or ADF/cofilin) - that sever and depolymerize ADP-actin filaments, thereby increasing the pool of available G-actin monomers. The structure of the network is further controlled by capping proteins that can bind to F-actin plus ends to terminate their growth, and thus limit the increase of free-growing plus ends. Finally, cross-linked structures are formed with the help of actin cross-linkers like filamin, and actin-bundling proteins like fascin, fimbrin and $\alpha$-actinin. $\alpha$-actinin and filamin are most present in lamellipodium structures, as fimbrin and fascin and most observed in filopodia \cite{Small:2002sw}. Finally, the same as well as other actin-binding proteins (like espin, fascin, fimbrin and villin) exist in other structures where actin-bundles are formed, like bristles, microvilli and stereocilia \cite{Revenu:2004fu}. In skeletal muscles, tropomyosin strengthens the actin filaments and prevents myosin motors from binding to actin when muscles need to be at rest\footnote{Up-to-date informations can be found at http://www.bms.ed.ac.uk/research/others/smaciver/Cyto-Topics/actinpage.htm.}. For a relatively recent review on the actin-binding proteins, see \cite{Remedios:2003jt}.

Microtubule-associated proteins (MAPs) have been classified into two types, and participate to microtubules' stability and organization. MAPs of Type I are large filamentous proteins that comprise a microtubule-binding domain and a projection domain, thereby controlling the spacing of microtubules. MAPs of Type II have similar structures and cross-link microtubules to membranes, intermediate filaments or other microtubules. In addition, both types of MAPs promote microtubule assembly and stability, and compete with motor proteins for binding sites, such that they participate in microtubule-transport regulation. Other MAPs that do not belong to these classes are denoted XMAPs, as they have been originally identified in the Xenopus-frog eggs. Among these are the plus-end-binding proteins (or +TIPs) that bind to the microtubule growing ends and participate in their stability, and the highly-conserved stathmin or oncoprotein 18 which, instead, destabilizes microtubules \cite{Andersen:2000db,Schuyler:2001hl}. The best understood microtubule end-binding proteins are the MCAKs (for mitotic centromere-associated kinesins), also known as Kin I kinesins, which are unusual kinesins in that, instead of moving along the surface of microtubules like other kinesin proteins do, they bind to microtubules' ends and trigger depolymerization in a processive way \cite{Desai:1999lr}. In particular, they depolymerize microtubules during mitosis to drive chromosome segregation \cite{Maney:2001fk}. For a review, see \cite{Howard:2003lo}.

\subsection{\label{Sse:CellAnchoring}Cell anchoring and regulatory pathways}

The two first steps of cell crawling in the Abercrombie classification consist in the protrusion of the leading edge and its adhesion to the substrate \cite{aber80,Lauffenburger:1996cr}. Although they were thought to be largely independent processes, evidences are accumulating that adhesion and protrusion are highly interrelated \cite{Ridley:2003wj,Bershadsky:2006px,Gupton:2006lr,Ananthakrishnan:2007kx}. Protrusion results primarily from actin polymerization at the leading edge of the migrating cell (see Section \ref{Se.Filaments}), and is regulated by the small GTPases Rho, Rac and Cdc42 \cite{Bishop:2000lr,Ridley:2001fk}. As Rho is known to activate actomyosin contractility, Rac and Cdc42 induce actin polymerization and the formation of actin-filled protrusions such as lamellipodia and filopodia \cite{Hall:2005qy}. Through these pathways, the cell can respond to the chemical composition of its environment, an example of chemotaxis: as a function of the gradients of chemoattractants or chemorepellants in its environment, the cell regulates its sites of fastest actin polymerization in order to move towards or away from the source\footnote{See the movies associated with ref \cite{Parent:1999yq}, as well as the one of a neutrophil cell that chases a bacterium at
http://www.biochemweb.org/fenteany/research/cell\_migration/movement\_movies.html.} \cite{Parent:1999yq}. Of primary importance for the accurate spatial regulation of these processes is the role of microtubules, which in some cell types play a crucial role in determining cell polarity and directional migration \cite{Kaverina:2002bs,Small:2002uq,Nelson:2003hs,Small:2003wo,Ridley:2003wj}. Microtubules have been proposed to activate Rac and Rho, the latest via the release of the GDP-GTP exchange factor GEF-H1 during microtubule depolymerization \cite{Krendel:2002kk,Waterman-Storer:1999lr}. 

\begin{figure}[h]
\scalebox{1}{
\includegraphics{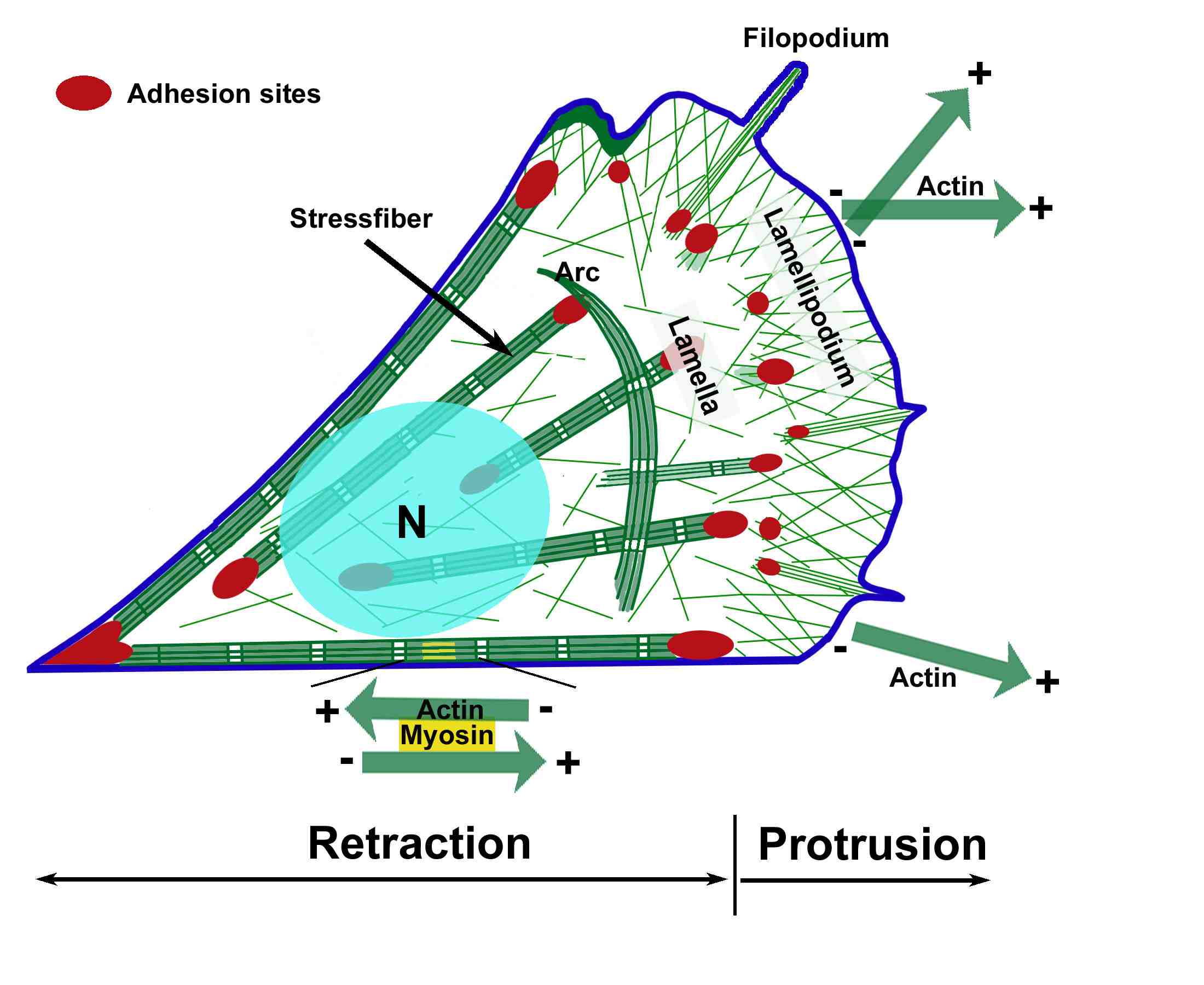}
}
\caption[cytoskeletonandcellmotility_fig11]
{\label{cytoskeletonandcellmotility_fig11}
Schematic representation of the actin cytoskeleton in a polarised fibroblast. The different organisational forms of actin filaments and their relations to adhesion sites to the substrate are depicted: diagonal actin filament meshwork in the lamellipodium, with associated radial bundles that sometimes protrude into filopodia; contractile bundles of actin (stress fibers) in the cell body and at the cell edge; and a loose actin network throughout the cell. Arc-shaped bundles are sometimes observed that move inwards under the dorsal cell surface (arc). The diagram shows an idealized cell: in reality, actin arrays are interconnected in various combinations and geometries. Adhesion sites are indicated in red. The flat region behind the lamellipodium and in front of the nucleus (N) is termed the lamella. At the cell front, in lamellipodia and filopodia, actin filaments are all polarized in one direction, with their fast-growing ends directed forward for producing pushing forces and inducing protrusion; in the cell body, actin filaments form bipolar assemblies with myosin proteins (stress fibers) for retraction. Source: courtesy of Vic Small; modified from ref. \cite{Kaverina:2002bs} with permission from Elsevier Limited.}
\end{figure}

\begin{figure}[h]
\scalebox{0.34}{
\includegraphics{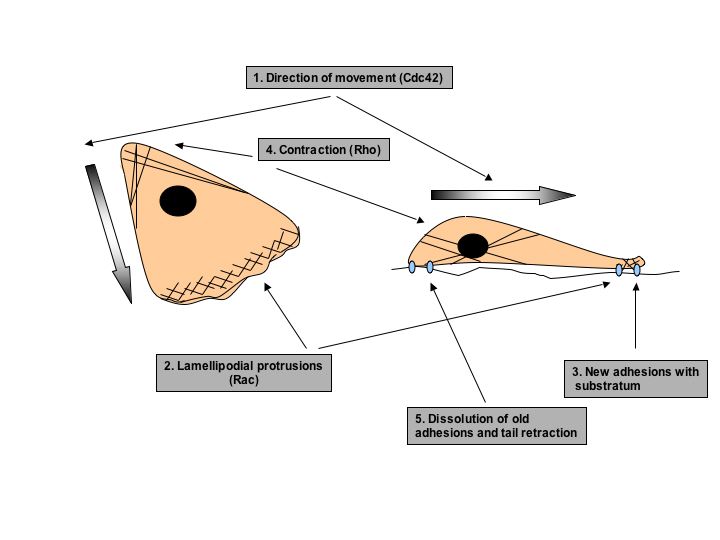}
}
\caption[cytoskeletonandcellmotility_fig12]
{\label{cytoskeletonandcellmotility_fig12}
Schematic representation of the integrated roles of Rho, Rac and Cdc42 proteins in regulating cell migration. By inducing actin-filament assembly, filipodia and focal-complexes formations, Cdc42 regulates the direction of migration (1). Rac induces actin polymerization at the cell periphery (2) and promotes lamellipodia protrusion. It also induces the formation of focal complexes at the leading edge (3). Rho plays a role in regulating longer-lived structures, namely activating actomyosin contraction in the stress fibers located in the cell body and at the rear (4), as well as promoting the assembly of focal-adhesion complexes. Source: courtesy of Alan Hall; reprinted from ref. \cite{Raftopoulou:2004lr} with permission from Elsevier Limited.}
\end{figure}

Adhesion occurs via the formation of adhesion sites, which rely primarily on molecules such as \emph{integrins}, also involved in regulating the cell behavior via different signal-transduction pathways \cite{Geiger:2001kl,Miranti:2002tg,Bershadsky:2006px}. An important aspect of that process is that it allows the cell to ``feel'' the mechanical properties of its environment. This has been shown to be important for the migration of fibroblasts, in that they seem to migrate preferentially towards regions of stiff substrates, a process referred to as \emph{durotaxis} \cite{Lo:2000cr}. In addition, the mechanical properties of the cell's environment have been proposed to be relevant for tissue-growth directionality, as well as cell differentiation \cite{Saez:2007rt,Engler:2006vn}. Adhesion sites can be roughly divided into two broad categories, namely \emph{focal complexes}, which locate beneath microspikes or filopodia, and \emph{focal adhesions}, which locate at the termini of stress fibers and serve in long-term anchorage \cite{Kaverina:2002bs} (See Fig. \ref{cytoskeletonandcellmotility_fig11}).
Interestingly, these adhesion sites are also regulated by small GTPases of the Rho, Rac and Cdc42 families: focal complexes are signaled via Rac1 and Cdc42, and can either turnover on a minute time-scale or differentiate into long-lived focal adhesions via the intervention of RhoA \cite{Kaverina:2002bs}. A schematic representation of the integrated role of the small GTPases in regulating cell migration can be seen in Fig. \ref{cytoskeletonandcellmotility_fig12}.
Such regulations are crucial for cell migration to occur optimally. Indeed, whereas adhesion sites are necessary at the leading edge of the cell to provide anchoring points on which the cell can exert traction forces, these need to be released at the rear for the cell to move forward. This results in a biphasic response of the cell-migration speed as a function of adhesive-ligand concentrations, in that too-low or too-high ligand concentrations prevent either the traction forces to be exerted, or the rear to be released \cite{Lauffenburger:1996cr,Jurado:2005wd}. How these regulatory pathways lead to a spatio-temporal feedback mechanism between actomyosin regulation and the focal-adhesion system is still under investigation \cite{Bershadsky:2006px,Gupton:2006lr}.

\subsection{\label{Sse:Prokaryotes}The prokaryotic cytoskeleton}

This section is completely independent of the rest of the article. Readers not interested in the biochemical composition of the prokaryotic cytoskeleton might want to skip this section.

As cytoskeletal protein's structures are highly conserved throughout the three domains of life (archaea, bacteria and eukarya), prokaryotic cytoskeletal proteins differ strongly in their sequences from their eukaryotic counterparts. For this reason, and the fact that prokaryotes have a relatively simple organization as compared to eukaryotic cells, it was long thought that they were lacking a cytoskeleton. It is only in the 1990s that prokaryotic homologs of tubulin, actin and intermediate filaments started to be discovered. The first bacterial cytoskeletal proteins to be brought to knowledge was the protein FtsZ, whose relation to tubulin was discovered independently by three groups in 1992 \cite{RayChaudhuri:1992ys,Boer:1992fr,Mukherjee:1993zr}. Later, it was found that FtsZ could assemble into protofilaments that can be either straight or curved as a function on the state of the nucleotides, similarly to microtubules \cite{Erickson:1996mz,Howard:2003lo}, and that its structure at the level of protein folding was nearly identical to that of tubulin \cite{Lowe:1998ly,Nogales:1998gf}. The second prokaryotic cytoskeletal proteins to be discovered were MreB and ParM also in 1992, and were shown to be distant relative of the actin superfamily by sophisticated sequence-alignment techniques \cite{Bork:1992ve}. Later, it is only in 2001 that MreB was proven to be capable of self-assembly into cytoskeletal filaments that resemble much closely F-actin structures \cite{Jones:2001ul,Ent:2001cr}. Finally, an homolog of intermediate filaments has been found recently in the bacterium \emph{Caulobacter crescentus}, but only in this particular species so far \cite{Ausmees:2003pd}. Since it is responsible for giving the bacterium its crescent shape, it was given the name of \emph{crescentin}.

Despite their sequencial differences with their eukaryotic counterparts, prokaryotic cytoskeletal proteins share with them strong homologies in their structural as well as functional properties \cite{Amos:2004bh,Michie:2006lq}. They are classified into four groups \cite{Shih:2006dq}: (i) Actin homologs are constituted by MreB and MreB homologs, ParM, and MamK. MreBs play an important role in a number of cellular functions, such as regulation of cell shape, chromosome segregation, establishment of cell polarity and organization of membranous organelles. ParM proteins are involved  essentially in plasmid partitioning, and MamK is involved in the subcellular organization of membrane-bounded organelles. Similarly to actin, MreB and ParM protein families present polymerization-depolymerization dynamics that are driven by ATP hydrolysis. Less is known about MamK. (ii) Tubulin homologs contain FtsZ and the BtubA/B proteins, which constitute two other families of GTPases as compared to tubulin. As FtsZ is crucially involved in cytokinesis via its ability to form contractile rings and spiral structures, the role of BtubA/B proteins, which are much less widespread in the bacterial kingdom, has less been characterized so far. (iii) The intermediate filaments' homolog crescentin has only been found in \emph{Caulobacter crescentus} and, as for its eukaryotic counterparts, is mainly involved in cell shape and structural integrity. (iv) Finally, the large MinD/ParA superfamily is made of prokaryotic cytoskeletal proteins that have no counterparts in eukaryotes. They however have the ability to organize into polymeric filaments, and present ATPase activity. Proteins of the MinD group are involved in placement of the bacterial and plasmid division sites, whereas proteins of the ParA subgroup are primarily involved in DNA partitioning.

Interestingly, cytoskeletal proteins seem to have been strongly conserved throughout evolution in each of the three separate domains of life, but differ quite substantially across domains \cite{Erickson:2007rr}. Bacterial FtsZs proteins are 40-50\% identical in sequence across species, and share even the same amount of similarities with their archaeal counterparts. Bacterial MreBs are generally 40\% conserved. Among eukaryotes, the conservation is even stronger: it reaches 75-85\% for tubulin and 88\% for actin, one of the most conserved protein in the eukaryotic domain. In the case of archaea, MreB and actin homologs have not yet been identified for sure \cite{Erickson:2007rr}.

\section{\label{Se.Filaments}Filament-Driven Motility}

Many kind of movements in eukaryotic cells are driven by polymerization-depolymerization mechanisms of cytoskeleton filaments, for which motor proteins \emph{per se} are not required. Instead, the chemical energy stored in high-energy hydrogen bounds (under the form of ATP or GTP) is converted into movement via treadmilling mechanisms \cite{Theriot:2000eu}. Two types of filaments have this ability, namely microtubules and actin filaments.

\subsection{Microtubule growth and catastrophes}

Microtubules are the stiffest of all cytoskeletal filaments, which confers them the ability to organize and stabilize both the cell structure and its transport network for internal communication and distribution. Depending on the cell need, they constantly reorganize or exert forces on the cell membrane or other organelles they transport. This is the case for example during the organization of the mitotic spindle, the structure formed prior to chromosomal segregation during mitosis \cite{Scholey:2003et}. There, chromosomes gather in a plane halfway from two microtubule-organizing centers - the centrosomes - that are located at each pole of the future dividing cell, and from which the microtubules tear the chromosome pairs apart \cite{Albertson:1984ht,Inoue:1995ve} (see Fig. \ref{cytoskeletonandcellmotility_fig6}, especially panels C to F). For this mechanism to happen, microtubules constantly exert pulling and pushing forces both on the chromosomes and the cell membrane, a mechanism that allows for correct positioning of the site of cell-division \cite{Yeh:2000az}. Coupled with kinesin-motor activity, correct positioning of the centrosomes relies crucially on the ability of microtubules to grow and shrink spontaneously, a dynamics that provides feedback to centrosome positioning and leads to oscillations orthogonal to the cell spindle axis \cite{Grill:2001hb,Grill:2005qv}. Such a mechanism is also responsible for the correct positioning of the nucleus in the fission yeast \emph{Schizosaccharomyces pompe} \cite{Tran:2001hw}.

Understanding microtubules' polymerization-depolymerization dynamics started with the first observation that they display phases of relatively slow growth, alternated with phases of rapid shrinkage \cite{Mitchison:1984ad}. Changes from one type of behavior to the other are referred to as \emph{catastrophes} for the conversion from growing to shrinking, and \emph{rescues} for the opposite transition. Observations of this behavior were further made in culture cells \cite{Sammak:1988ov} and cellular extracts \cite{Belmont:1990xr}, which confirmed the existence of such dynamic instabilities \emph{in vivo}. During mitotic-spindle formation, it has later been shown that the specialized structures that connect the microtubules to the chromosomes, the \emph{kinetochores}, can ``capture'' and stabilize growing microtubules, preventing them from undergoing catastrophes \cite{Hayden:1990qw}. For a review, see \cite{Karsenti:2001vz}.

Further characterization of microtubules' biomechanical properties came from experimental studies of the forces produced by their polymerization-based growth. Analyzing force-induced microtubule buckling \cite{Dogterom:1997kl}, microtubule forces were characterized as being potentially as high as those produced by motor proteins - typically a few pico-Newtons \cite{Janson:2004lp} - and to be able to deform membranes \cite{PhysRevLett.79.4497} or center asters in mirofabricated chambers \cite{Holy:1997fz,Faivre-Moskalenko:2002cv}, a mechanism that imitates nucleus positioning in fission yeast. For reviews, see \cite{Howard:2003lo,Dogterom:2005vl}.

\subsection{\label{Sse.ActinGels}Actin gels}

As earlier stated, the first step of amoeboid motility in the original Abercrombie classification occurs via protrusion of the leading edge of the cell. This mechanism relies mainly on the polymerization dynamics of actin filaments \cite{Pantaloni:2001dp,Pollard:2003fh,Mogilner:2006dz}. Actin polymerization is known to play a primary role at the plasma membrane, where it is nucleated by proteins of the WASP family via Arp2/3 complexes (see Section \ref{Sse:OtherProteins}). It has also been proposed to be responsible for driving endocytosis and the movement of endosomes, both in cultured cells and yeast \cite{Merrifield:1999ht,Kaksonen:2003xh}.

\begin{figure}[h]
\scalebox{0.3}{
\includegraphics{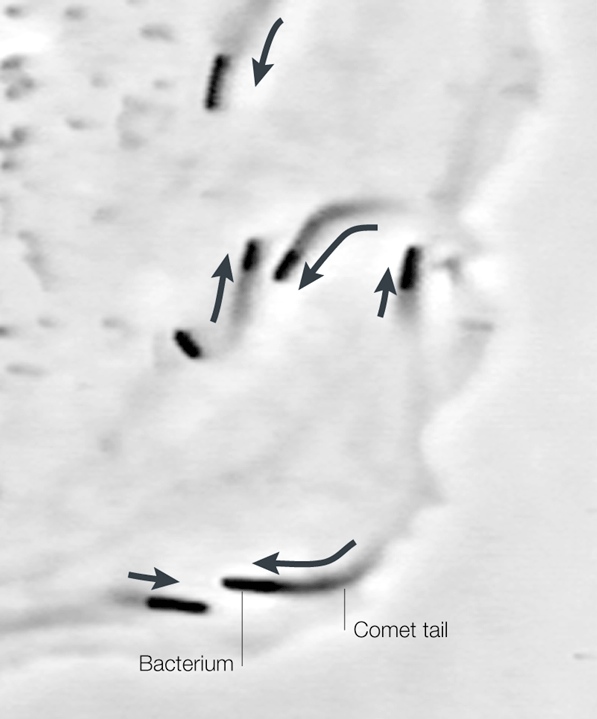}
}
\caption[cytoskeletonandcellmotility_fig13]
{\label{cytoskeletonandcellmotility_fig13}
\emph{Listeria} bacteria moving in a cell via actin-based motility. This is a phase-contrast micrograph, a single frame from a video sequence\footnote{The whole movie can be seen on the following webpage: http://cmgm.stanford.edu/theriot/movies.htm.}. The kidney epithelial cell was infected about five hours before the acquisition of this video sequence. All of the bacteria in this cell are clonal descendants of a single individual. A bacterium and its associated comet tail are labelled. Bacteria are moving in the direction of the arrows. Source: reprinted from ref. \cite{Cameron:2000rr} with permission from Nature Publishing Group.}
\end{figure}

Our understanding of eukaryotic actin-based motility has grandly benefitted from the motility mechanism of the bacterium \emph{Listeria monocytogenes}. This pathogene moves at velocities of the order of several micro-meters per minute by nucleating the formation of an actin ``comet-tail'' that, while polymerizing thanks the host's cytoskeletal machinery, pushes the pathogene forward \cite{Theriot:1992vl} (see Fig. \ref{cytoskeletonandcellmotility_fig13}).
This particular motility mechanism, studied in \emph{in vitro} assays, has allowed for the identification of the minimal set of proteins needed to actin-based motility, as well as the role of several of the main actin-related proteins \cite{Welch:1998kh,Loisel:1999lh}. It has also been used as a probe for the cell cytoskeleton network structures and visco-elastic properties in a position-dependent manner \cite{Lacayo:2004xd}, and has shed light into the basic elementary principles of actin-based motility \cite{Pantaloni:2001dp,Frischknecht:2001wm,Gruenheid:2003zv,Mogilner:2006dz}.

Except for very recent reports \cite{Footer:2007wd}, nearly no force measurement has yet been done on single actin filaments. Due to their smaller bending rigidity, the corresponding stall force is expected to be orders of magnitude smaller than that of microtubules because of buckling phenomena. Instead, large forces can only be obtained when highly cross-linked actin filaments work as a whole and form a relatively rigid network, as it is the case in filopodia protrusion. Forces generated during actin-based propulsion have been measured on polymerizing actin gels, in particular using \emph{in vitro} assays based on artificial biomimetic systems. Forces in the range of a few nano-Newtons have been found for gel comets originating from 2-$\mu$m-size polystyrene beads \cite{Marcy:2004bs}.

Other bio-mechanical characterizations of actin network's properties concern the study of its gel-like viscoelastic properties. In particular, transitions between a solid-like elastic material and a solution-like viscous material have been observed \cite{Fukui:2002ad}. These could rely in part on the biochemical-dependent mechanical properties of the actin filaments themselves \cite{Isambert:1995fp}, on the generic properties of such semiflexible-filament networks \cite{Gardel:2004os}, or on the activity of motor proteins that help disentangling the network and thereby lead to its fluidization \cite{Humphrey:2002mw}. It has also been observed that cross-linked actin networks display an increase of their elastic modulus as a function of the stress applied, a nonlinear behavior known as \emph{stress-stiffening} \cite{Storm:2005cz,Fernandez:2006cl}. This might explain partly the interestingly rich properties of cellular rheology \cite{Hoffman:2006xq,Kasza:2007tx}, which have been partly reproduced in \emph{in vitro} measurements \cite{Gardel:2006qq}. Among these, dynamical scaling of the stress stiffening (see e.g. \cite{Gardel:2004tw}) has been proposed to be the signature of underlying self-similar mechanical properties of the cell cytoskeleton \cite{Fabry:2001ek,Balland:2006oa}. Finally, the intermediate filaments as well as the biochemical environment or preparation of the actin network have been proposed to play an important role in modulating its rheological properties \cite{Janmey:1991mz,Gardel:2004os}. This might contribute to the observed local changes in the elasticity of the cell as it moves, a crucial aspect for driving its motility \cite{Fukui:2002ad,Ananthakrishnan:2007kx}.

\subsection{Modeling polymerization forces}

With general thermodynamic considerations, growth velocities of polymerizing filaments can be understood as follows: if $k_{\rm{on}}$ and $k_{\rm{off}}$ are the association and dissociation constants for monomers at the polymer tip, and $\delta$ is the distance a filament grows under addition of a single monomer, a typical growth velocity of the polymerizing filament is given by: $v=\delta\left[ k_{\rm{on}} - k_{\rm{off}} \right]$. When experiencing a force $f$ opposing polymerization, like the cellular membrane resistance at the leading edge of the advancing cell, filament-growth velocity becomes:
\begin{equation}\label{Eq.velocity1}
v(f)=\delta \left[ k_{\rm{on}}\exp{\left(-q\frac{f a_1}{k_B T}\right)}-k_{\rm{off}}\exp{\left( (1-q)\frac{f a_1}{k_B T}\right)} \right].
\end{equation}
In this expression, $k_B T$ represents thermal energy, $f a_1$ is the most probable work needed to add a monomer in the presence of the force $f$, and $q$ is a parameter describing how much the force $f$ influences the on-rate as compared to the off-rate. Under these assumptions, the maximal force a given filament can produce via polymerization, or \emph{stall force}, is expressed as $f_s=(k_B T/a_1)\ln{\left( k_{\rm{on}}/k_{\rm{off}}\right)}$. Even though good overall agreement with experimental data was obtained for individual microtubules while choosing  $13\, a_1=a$ and $q=1$ (with $a$ being the size of a tubulin monomer) \cite{Dogterom:1997kl}, the so-derived stall force was too large as compared with experimental measurements. This led to revising the dynamics of the microtubule-polymerizing end, proposing $a_1\simeq a$ and $q\simeq0.22$ as better fitting parameters, pointing to a rich dynamics of microtubule polymerization \cite{Kolomeisky:2001rw}.

To understand the origin of polymerization forces, the standard microscopic model relies on the ratchet mechanism, a rectified Brownian motion originally introduced in this context by Peskin \emph{et al.} \cite{Peskin:1993gf} to explain filopodia protrusion, \emph{Listeria} propulsion as well as protein translocation. Filopodia protrusion in particular is thought to rely essentially on actin-polymerization forces: when reaching the cell membrane, growing F-actin filaments feel a force opposing their growth, and therefore exert a force on the membrane. Because of thermal fluctuations and membrane's as well as actin-filaments' finite bending rigidities, some space is constantly opened between the growing filament and the membrane. From time to time, an additional monomer can thereby be added to the growing filament, which pushes the membrane forward\footnote{For an animated illustration, see http://www.jhu.edu/cmml/movies/anim/eBRatchet2.swf.}. In the simplest case of a single stiff protofilament, the distribution $P$ of distances $x$ between the filament and the membrane is given by the following Fokker-Planck equation:
\begin{widetext}
\begin{eqnarray}
\partial_t P(x) &=& D\,\partial_x^2 P(x) + f \frac{D}{k_B T}\,\partial_x P(x) + k_{\rm{on}} P(x+\delta) - k_{\rm{off}} P(x) \quad \textrm{for} \quad x<\delta,\nonumber\\
\partial_t P(x) &=& D\,\partial_x^2 P(x) + f \frac{D}{k_B T}\,\partial_x P(x) + k_{\rm{on}} [P(x+\delta) - P(x)] - k_{\rm{off}} [P(x) - P(x-\delta)] \quad \textrm{for} \quad x>\delta,
\end{eqnarray}
\end{widetext}
where notations are similar to the ones used in Eq. (\ref{Eq.velocity1}), and to which must be added the effective diffusion coefficient $D$ for the distances $x$ between the filament and the membrane. The time-dependence of $P$ is implicit. Using vanishing-current conditions at the leading edge $x=0$, the stall force can be obtained and is given by an expression analog to that of microtubules models, with $a_1=\delta$ being the size of a G-actin monomer. Including bending fluctuations of the growing filament, this led to the Elastic Brownian Ratchet Model \cite{Mogilner:1996qf}, a generalization of which is the Tethered Elastic Brownian Ratchet Model \cite{Mogilner:2003rc} that considers that some filaments are attached to the membrane via protein complexes (as it has been observed in the \emph{Listeria}-propulsion mechanism for example) and therefore do not exert polymerization forces. When typical parameter values are plugged into these models, single actin-filament force generation is estimated to be of the order of 5-7 pN \cite{Mogilner:2003qd}. Taken into account that at the leading edge several hundreds of actin filaments per micron work together to drive the cell forward, the resulting force is of the order of nanonewtons per micron \cite{Mogilner:2003rc}, a force large enough to tackle the membrane load and resistance. However, it has since then been claimed that motor proteins, called \emph{end-tracking motors}, should be required to explain observed forces in the case of \emph{Listeria} propulsion for example \cite{Dickinson:2004sp}. This work has been reviewed in \cite{Mogilner:2006dz}.

Lateral interactions between filaments in an actin network have been investigated via models that take into account the branching structure of the network \cite{Carlsson:2001tg,Carlsson:2003hc}. In particular, Autocatalytic Models assume that new branches are generated from existing ones, which leads to a growth velocity that is independent of the load \cite{Carlsson:2003hc}. To investigate the consequences of these models, two approaches have been followed, namely stochastic simulations of the growing actin network, tracking each filament position and orientation \cite{Carlsson:2001tg}, and deterministic rate equations that include growth, capping and branching rates, and which led to a comparison between ratchet and autocalytic models \cite{Carlsson:2003hc}. Experimental tests of the two models have been performed in \emph{in-vitro} systems, using \emph{Listeria} propulsion as well biomimetic systems \cite{Upadhyaya:2003ov} (see next paragraph). While some \emph{Listeria} studies favored the Tethered Elastic Brownian Ratchet Model \cite{McGrath:2003mi}, some studies using biomimetic systems favored the Autocatalytic Model \cite{Wiesner:2003eq}, and several others neither of them. A possible explanation for these apparently contradictory results may be that different experimental studies led to analyzing different regimes of the force-velocity curve.

\subsection{A model system for studying actin-based motility: The bacterium \emph{Listeria monocytogenes}}

As earlier stated, our understanding of eukaryotic actin-based motility has grandly benefited from the motility mechanism of the bacterium \emph{Listeria monocytogenes} (see Section \ref{Sse.ActinGels}). While velocities of \emph{Listeria} bacteria in a homogeneous environment are typically constant, some mutants progress in a saltatory manner \cite{Lasa:1997ee}. This observation has been later reproduced in \emph{in vitro} motility assays using latex beads coated with the bacterium transmembrane protein ActA (that further recruits Arp2/3) \cite{Cameron:1999bl}, or directly with VCA proteins, a sub-domain of WASP that is responsible for actin-branching and polymerization nucleation \cite{Bernheim-Groswasser:2002uf}. Such biomimetic systems have allowed for the direct measurement of the characteristic polymerization force that is produced by an actin gel \cite{Marcy:2004bs}, and for the overall study of actin-based motility mechanisms in simple and well-controlled conditions \cite{Upadhyaya:2003ov,Plastino:2005yq,Mogilner:2006dz}.

Theoretical understanding of such actin-based propulsion mechanisms has come from two different angles, namely molecular and mesoscopic, continuum models. We have already reviewed the molecular models that rely on brownian-ratchet mechanisms. They, in particular, have led to force-velocity curves that are consistent with some observations of \emph{Listeria} motion \cite{Mogilner:2003rc,McGrath:2003mi}. Continuum models describe the actin network as a compressible elastic gel with an elastic modulus of about 5000 Pa \cite{Noireaux:2000hc,Gerbal:2000hw,Marcy:2004bs} (see also Section \ref{Sse.Gels}). When growing over a curved surface like the bacterium \emph{Listeria} or a coated bead, the gel deforms as it grows by monomer additions on the particle surface, which in turn generates a stress that pushes the particle forward \cite{Noireaux:2000hc,Gerbal:2000hw}. Monomer transport to the inner surface of the growing gel is purely diffusive, with a diffusion constant that has been estimated to be of the order of 2 $\mu \rm{m}^2$/s for actin monomers in an ActA-produced gel \cite{Plastino:2004ku}. When originally initiated on a spherical object like a rigid bead, the growth of the gel layer starts isotropic, but ruptures into a comet-type growth because of mechanical instability. The instability relies on a positive feedback that involves creation of a tensile stress as the gel grows because of geometrical effects, and enhancement of the depolymerization rate or rupture of the gel in regions of enhanced tensile stress \cite{Sekimoto:2004wv}. The instability occurs less rapidly with increasing bead size, which explains why movement is more often observed with small beads. This mechanism of tensile-stress accumulation and rupture can also explain the saltatory motion observed with \emph{Listeria} mutants and coated beads in some conditions \cite{Bernheim-Groswasser:2002uf}: rapid phases of motion are due to the sudden rupture of the gel that pushes the particle forward, as slow phases of motion correspond to progressive build-up of lateral tensile stress. Depending on the size of the bead as well as the concentration of proteins at its surface, this dynamic instability can be present or not, which explains the observation of both continuous and saltatory regimes with coated beads as well as \emph{Listeria} bacteria \cite{Bernheim-Groswasser:2002uf,Bernheim-Groswasser:2005oh}.

\begin{figure}[h]
\scalebox{0.13}{
\includegraphics{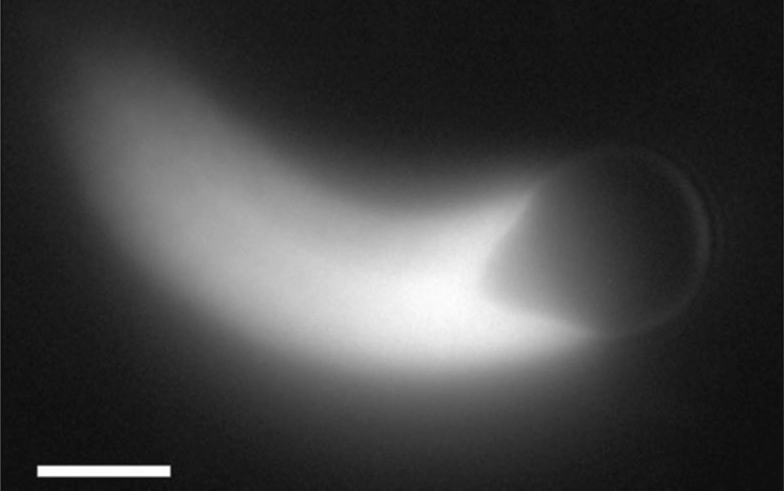}
}
\scalebox{0.155}{
\includegraphics{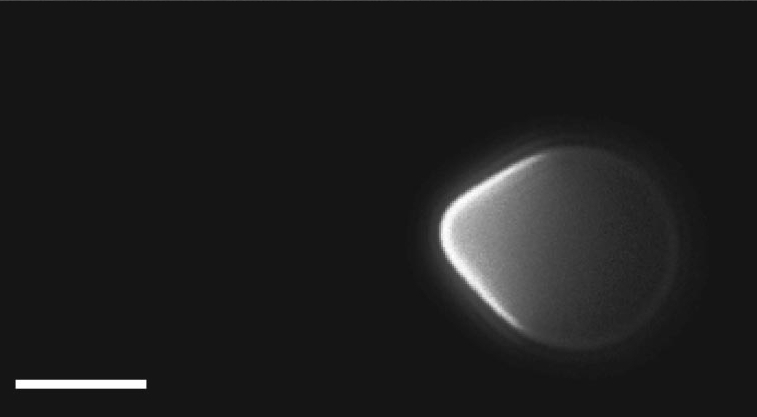}
}
\caption[cytoskeletonandcellmotility_fig14]
{\label{cytoskeletonandcellmotility_fig14}
Actin-based propulsion with liquid drops. Oil drops are covered with VCA, placed in cell extracts that are supplemented with actin, and observed by fluorescence microscopy. Left panel: Note the bright actin comet and the pear-like shape of the droplet due to squeezing forces exerted by the actin gel. Scale bar is 4 $\mu$m. Right panel: VCA is  labeled with fluorescin isothiocyanate (FITC). Note the inhomogeneous distribution of the actin-polymerization promoter on the surface of the droplet. Scale bar is 3 $\mu$m. Source: reprinted from ref. \cite{Boukellal:2004qt} with permission from The American Physical Society.}
\end{figure}

To further explore the properties of the actin gel and the \emph{Listeria}-propulsion mechanism, experiments with soft objects like liposomes \cite{Upadhyaya:2003qz,Giardini:2003ff}, endosomes \cite{Taunton:2000bf} and oil droplets \cite{Boukellal:2004qt} have been performed. They show that the actin gel squeezes the object, compressing its sides and pulling its rear, an effect that gives it a pear-like shape (see Fig. \ref{cytoskeletonandcellmotility_fig14}).
Analysis of the contour of the deformed objects provides informations on the distribution of the normal stress on the surface of the object. This could in principle allow for the derivation of the total force exerted on the load in the case for example of oil droplets, where interfacial tension is measurable and normal stress can be deduced from Laplace's law \cite{Boukellal:2004qt}. But in fact, assuming a constant surface tension, the integration of the normal stress over the surface of the droplet gives a zero net value of the force that is independent of the droplet shape, the latest being regulated by the variation of the polymerization velocity with normal stress. Instead, the distribution of actin-polymerization promoters on the surface of the droplet follows the gel elastic deformations, which in turn creates pressure variations inside the droplet, and thereby surface-tension gradients along its surface (see Fig. \ref{cytoskeletonandcellmotility_fig14}); and these are at the origin of the final non-zero net pushing force \cite{Boukellal:2004qt}. Finally, direct observation of the actin comet during its growth on coated beads has shown that the actin gel constantly undergoes deformations that depend on the protein composition of the motility medium they are placed in \cite{Paluch:2006tz}. As a function of bead size and the concentration of cross-linkers or regulatory proteins, the bead velocity can be limited either by diffusion of the monomers to the coated surface, by polymerization velocity at the surface of the bead, or by the elastic stress built up in the gel. These findings, supported by experimental results, buttress the idea that actin-based movement is governed by the mechanical properties of the actin network, themselves tightly regulated by the proteins that are involved in actin dynamics and assembly (see Section \ref{Sse:OtherProteins}).

\subsection{\label{Sse.MSP}Another example of filament-driven amoeboid motility: The nematode sperm cell}

Even though, as described above, protrusion of the leading edge in amoeboid motility is most commonly actin-driven, other cells, the nematode sperm cells, use another cytoskeletal protein to drive their motility: the Major Sperm Protein (MSP) \cite{Roberts:2000wd}. Nematodes constitute one of the most common phyla of all animal kingdom, with over 80.000 different described species, and their sperm is thought to be the only eukaryote cell type that do not possess the globular protein G-actin. These cells offer an ideal to study cell crawling since, dislike actin, MSP is a simpler, more specialized protein that do not possess as many regulatory or associated proteins, and in particular is not known to bind any molecular motor (at the exclusion of end-tracking proteins). MSPs being also apolar \cite{Bullock:1996yf}, the nematode motility constitutes one of the simplest of all cytoskeleton-driven motility mechanisms known to operate \emph{in vivo} \cite{Roberts:2000wd,Wolgemuth:2005nx}.

Similarly to what has been done with biomimetic systems to study actin-based propulsion, motility assays using vesicles derived from the leading-edge of nematode sperm cells of \emph{Ascaris} species, have shed light into the mechanisms at play \cite{Italiano:1996mi}. In the presence of ATP, growth of MSP fibers are capable of pushing the vesicle forward, their polymerization being driven by specialized proteins located within the vesicle membrane, a mechanism that ressembles very much the \emph{Listeria}-propulsion mechanism. But, contrary to ATP-driven actin treadmilling, MSPs assemble into apolar filaments and lack a nucleotide binding site for ATP hydrolysis. To power membrane protrusion, it has recently been proposed that motor end-tracking proteins processively polymerize MSP filaments, while keeping the elongating filaments' ends in contact with membrane-associated proteins \cite{Dickinson:2007xp}. For cell progression however, a second force is required, namely a traction force that pulls the cell body forward once the advancing lamellipodium has been anchored to the substrate on which the cell is crawling. In actin-based amoeboid motility, this process is motor-driven, but nematode sperm cells use instead the sensitivity of their MSP to pH, whose decrease provokes reorganization, depolymerization and \emph{in fine} contraction of the network \cite{Italiano:1999eq,Miao:2003eu,Wolgemuth:2005nx}. Polarity in the cell is maintained by an influx of protons close to the cell body, which creates a pH gradient in the lamellipodium and powers this process \cite{King:1994cq}.

To quantitatively understand the mechanism underlying this motility, both microscopic and phenomenological models have been proposed. In the proposed microscopic models, mechanisms underlying the traction-force generation by solely cytoskeletal disassembly can be qualitatively understood as follows \cite{Bottino:2002oq,Wolgemuth:2005nx}: because of pH gradient, MSP filaments tend to bundle at the front, and split apart and disassemble at the rear \cite{Miao:2003eu}. A bundle with $N$ filaments being much stiffer than $N$ isolated individuals (with an effective persistence length of $N^2 L_p$ as each individual has a persistence length $L_p$), it pushes the cell membrane at the leading edge where filaments are bundled, while splitting filaments exert contractile forces at the rear. Indeed, because of entropic effects, filaments tend to retract once split apart. Finally, even further decrease in pH creates weakening of the attachments and dissociation of the filaments for monomeric MSPs to be recycled at the front \cite{Wolgemuth:2005nx} (see also \cite{Mogilner:2003qr}). In the proposed phenomenological approach however \cite{Joanny:2003ij}, the sensitivity to pH is described as influencing the equilibrium swelling properties of the gel only. As the gel treadmills towards the rear end where acidic conditions are found, it tends to contract by an isotropic multiplicative factor $\Lambda$ that is position-dependent. General elasticity theory of continuous media allows to express the strain tensor as:
\begin{equation}
u_{\alpha\beta}=\frac{1}{2}(1-\Lambda^2)\delta_{\alpha\beta}+\frac{1}{2}(\partial_{\alpha}u_{\beta}+\partial_{\beta}u_{\alpha}),
\end{equation}
where $u_{\alpha}$ are the components of the displacement vector (with $\alpha=x,z$). Assuming linear elasticity theory\footnote{For an introduction to the elasticity of continuous media, see, e.g., \cite{LandauLifschitzElasticity}.}, the stress tensor is then obtained as $\sigma_{\alpha\beta}=\lambda u_{\gamma\gamma}\delta_{\alpha\beta}+2\mu u_{\alpha\beta}$, where $\lambda$ and $\mu$ are the Lam\'e coefficients, which further leads to a position-dependent tensile stress as was introduced phenomenologically in \cite{Bottino:2002oq}. While traveling through the lamellipodium, tangential stress builds up, which leads to rupture of the adhesion points once a critical force has been passed, and eventually drags the cell body forward. Therefore, within this framework, only one parameter is directly controlled by the pH - namely $\Lambda$ - and the pH in particular does not need to influence directly adhesion strength.

\section{\label{Se.Motors}Motor-Driven Motility}

\subsection{Generic considerations}

Despite the major role played by polymerization forces in cellular motility, and in particular as we have previously seen in amoeboid motility, a vast amount of diverse motile processes in eukaryotic cells is driven by motor proteins (see Section \ref{Sse.Motors}). Theoretical studies of molecular motors started with the cross-bridge model published independently by A. F. Huxley and H. E. Huxley to explain the relative sliding of myosin filaments with respect to actin filaments in cross-striated muscle fibers \cite{HUXLEY:1957ys,HUXLEY:1957rt}. This approach was later formalized by Hill \cite{Hill:1974bh}, who introduced the notion of different ``states'' of a motor protein, each of these corresponding to a thermodynamic-equilibrium state. Interpretation of these different states was given in terms of different conformations of the motor protein and its interaction with the filament, or in terms of the state of the hydrolysis reaction of ATP, or both \cite{Leibler:1993ph,Duke:1996sf}. Justification for considering different thermodynamic-equilibrium states relied on the observation that for the transient response of muscles, the fastest response was known to be in the range of milliseconds, as thermal equilibrium on molecular characteristic length scales of 10 nm occurs after at most a few hundreds of nanoseconds. In this class of models, progression of the motor along the filament relies on asymmetric transition rates of the particle between the different states, for which asymmetry of the filament and energy consumption by the motor is required. Typically, after one cycle of conformational states, the motor protein has progressed by one or several allowed binding sites on the periodic lattice represented by the cytoskeletal filament. In between, up to five or six different states could be involved \cite{Lymn:1971dn,Hill:1974bh}. Experimental confirmation came later with the direct observation of walking steps displayed by advancing molecular motors \cite{Spudich:1994ud}. Such observations were first obtained studying kinesin motors in \emph{in-vitro} motility assays, and later with myosin motors displacing a filament that was attached to two glass beads placed in laser-trap potentials \cite{Bustamante:2000dq}. For a review, see \cite{Schliwa:2003yb}.

Another class of models relies on the generalization of Feynman's famous ``thermal ratchet'', in which the presence of different heat baths (namely thermal baths at different temperatures) can rectify the brownian motion of a given particle and lead to its directed motion \cite{feyn63}. For motor proteins, as we have already discussed, temperature inhomogoneities in the system cannot hold long enough to ground the mechanism. Instead, various different isothermal rectifying models have been discussed to describe the underlying mechanisms of different biophysical processes \cite{Magnasco:1993yi,Astumian:1997pd}. Among these, one can mention the translocation of proteins and force-generation by linear molecular motors (which includes cytoskeletal motors, but also motors acting on DNA or RNA, like DNA-polymerases, RNA-polymerases and helicases), the ion transport in ion pumps, and the rotary-motor processes such as the one found in the F0F1-ATPase or the bacterial flagellar motor. Such isothermal rectifying processes and their underlying physical principles have been extensively reviewed in \cite{juli97b,reim02}. They all rely on a Langevin type of description of an overdamped particle of position $x$, moving in a spatially-periodic potential $W(x)$ that reflects the motor-filament interaction, and subjected to a viscous friction with coefficient $\xi$ and a fluctuating force $f(t)$ that reflects the stochasticity of thermal fluctuations:\\
\begin{equation}
\xi\,\frac{dx}{dt}=-\partial_x W(x)+f(t).
\end{equation}
To rectify brownian motion, three different approaches have been mainly followed, namely  (i) random forces $f(t)$ whose fluctuations do not satisfy the fluctuation-dissipation (FD) relation, (ii) fluctuating potentials $W(x,t)$ that are time-dependent, and (iii) particle fluctuating between states, where different states indexed by $i=1,...,N$ reflects different conformations of the protein and interactions with the filament.

In the following, no attempt will be made to extensively present the literature on molecular motors. We shall instead only briefly sketch the generic considerations of the main proposed models, and focus more closely on a particular example of them, the two-state model, which has allowed for an understanding of the appearance of spontaneous oscillations in systems of coupled motors. This generic mechanism has been proposed to underly axonemal beating, the generic mechanism that powers eukaryotic flagellar and ciliary-based motilities.

\subsection{Phenomenological description close to thermodynamic equilibrium}

Sufficiently close to thermal equilibrium, out-of-equilibrium perturbations can be described using a generic linear-response theory that introduces generalized forces which drive generalized currents \cite{deGrootMazurB}. In the context of molecular motors, the generalized forces that drive the system out of equilibrium are the mechanical force $f_{\rm{ext}}$ acting on the motor (including drag), and the chemical-potential difference $\Delta\mu$ of the chemical reaction ATP$\rightleftharpoons$ADP+$\rm{P}_{\rm{i}}$ that drives motor power \cite{juli97b,Parmeggiani:1999fq}. Linear-response theory then gives:
\begin{eqnarray}
v&=&\lambda_{11}f_{\rm{ext}}+\lambda_{12}\Delta\mu,\nonumber\\
r&=&\lambda_{21}f_{\rm{ext}}+\lambda_{22}\Delta\mu,
\end{eqnarray}
where the coefficients $\lambda_{ij}$ are phenomenological response coefficients. Here $\lambda_{11}$ and $\lambda_{22}$ can be viewed respectively as a standard and generalized mobilities, and $\lambda_{12}$ and $\lambda_{21}$ as mechano-chemical couplings. Onsager relations impose that $\lambda_{12}=\lambda_{21}$, and the Second Law of Thermodynamics insures that the dissipation rate is positive: $T\dot{S}=f_{\rm{ext}}v+r\Delta\mu\geq 0$. Whenever both of the two terms that appear in this inequality are positive, the system is passive, but it works as a motor when $f_{\rm{ext}}v<0$, and as generator of chemical energy when $r\Delta\mu<0$. The latter function is not known for linear motors, but is the common mode of operation of F0F1-ATPase, the protein complex that synthetizes ATP from electro-chemical energy that is stored in proton gradients \cite{cellB,Yoshida:2001kl} (see Section \ref{Sse.Motors}). The reversibility of this rotary engine can be related to the predicted reversibility that comes out of linear-response theory: in the absence of an external force, reversing the chemical potential difference $\Delta\mu$ should reverse the sign of the velocity $v$ without a need for a change in the mechanism.

\subsection{Hopping and transport models}

Within the first class of models that we have mentioned earlier, namely hopping models between different discrete equilibrium states of the motor-filament system, generic transition rates and periodicity in theses transition rates, related to periodicity of the filament, are generally assumed. Within this framework, one can calculate the mean velocity $v$ and the diffusion coefficient $D$ of the molecular motor from analyzing the generic associated Master Equation \cite{derr83}. For non-zero mean velocity to occur, at least one of the transitions between states must break detailed balance, a feature that can be associated with chemical energy consumption. In the simplest case of only two possible states of the motor protein, one can derive simple compact expressions for $v$ and $D$ \cite{Fisher:1999kl}. Their dependence on the external force further leads to the derivation of the force-velocity curve, as well as a simple expression for the stall force, namely the force at which the motor protein ceases to progress on average.

To describe protein trafficking on a filament where many motors are simultaneously engaged, like it is commonly the case for example in organelle transport by kinesin proteins along microtubules, one can reduce the number of states that a motor can occupy to one per filament binding site. Motors are then represented by particles that move on a one-dimensional lattice with homogeneous transition rates, to which attachment and detachment rates from and toward the bulk can be added. This description belongs to a class of driven lattice-gas models that are used to study various transport phenomena, like ionic transport in solids or traffic flow with bulk on-off ramps \cite{schmittB95}. In the simplest case of the absence of particle attachment and detachment, the model reduces to the Asymmetric Simple Exclusion Process (ASEP) \cite{schutzB01}, originally introduced to describe the translation of messenger RNA by ribosomes \cite{MacDonald:1968xr}. Including attachment-detachment rates, the next simplest case describes the space surrounding the filament as a reservoir of uniformly-distributed particles \cite{Kruse:2002rz,Parmeggiani:2003bf}. A second possibility is to include the dynamics of unbound particles explicitly, for example on a cubic lattice \cite{Lipowsky:2001ph}. Boundary terms can also play an important role, and different possible choices have been considered depending on the biological situation \cite{Klumpp:2004gy}.

Consequences of these models are illustrated by various important phenomena. Among these, one can find the followings: anomalous transport due to repeated attachments and detachments \cite{AJDARI:1995qy,Lipowsky:2001ph,Nieuwenhuizen:2004sh}, domain walls that separate regions of high and low motor densities in the filament \cite{Lipowsky:2001ph,Parmeggiani:2003bf}, phase separation in systems with two motor species \cite{Evans:1995wy}, and phase transitions when cooperative binding-unbinding is introduced \cite{klum04b}. For a recent review on these collective traffic phenomena, see \cite{chow07} and references therein.

\subsection{The two-state model}

\begin{figure}[h]
\scalebox{0.1}{
\includegraphics{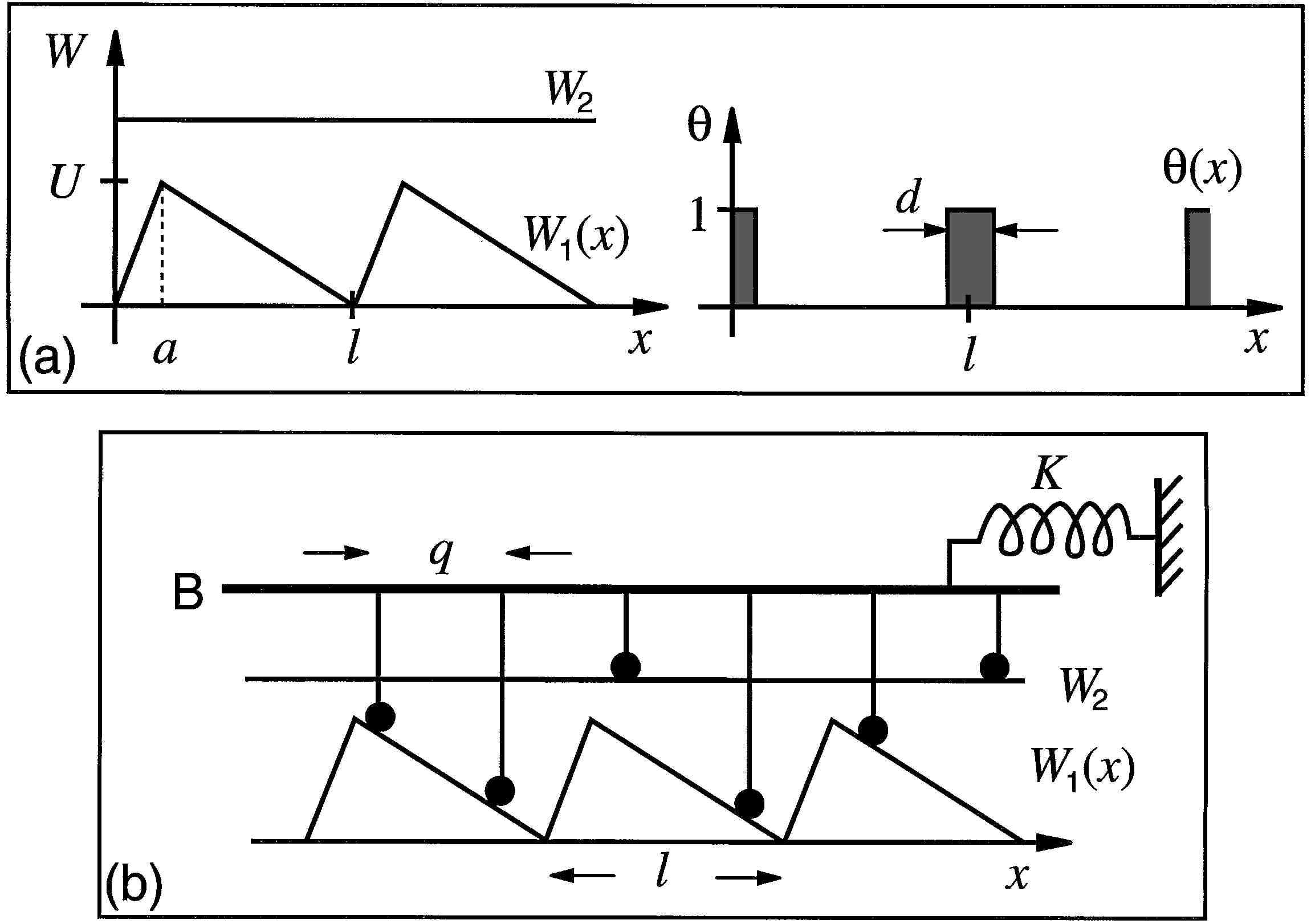}
}
\caption[cytoskeletonandcellmotility_fig15]
{\label{cytoskeletonandcellmotility_fig15}
Schematic representation of the two-state model as used in ref. \cite{juli97PhysRevLett.78.4510} for the calculation of the collective behavior of rigidly-coupled non-processive motors. (a), left panel: $W_1$ represents the attached state of the motor to the filament and is therefore asymmetric and $l$-periodic; $W_2$ represents the detached state where no interaction with the filament occurs. (a), right panel: Illustration of the $l$-periodic function $\Theta(x)$ as it appears in Eq. (\ref{Eq.BreakingDetailedBalance}) and which represents ATP consumption in the system, here at preferred locations of the motor on the filament, or at preferred configurations of the motor protein for ATP hydrolysis. (b) Schematic representation of a collection of rigidly-coupled motors with periodicity $q$ interacting with the filament, and coupled to its framework by a spring of stiffness $K$. Under some conditions, this model leads to spontaneous oscillations. Both structures and behaviors are reminiscent of skeletal-muscle myofibrils' oscillations or axonemal beating (see section \ref{SSe.AxonemalBeating}). Source: reprinted from ref. \cite{juli97PhysRevLett.78.4510} with permission from The American Physical Society.}
\end{figure}

One model that proved to be particularly useful for describing the rectification of brownian motion via coupling to chemical hydrolysis reactions, is the so-called ``two-state model''. In this description, the molecular motor switches stochastically between two different interaction states with the filament, that are described by two different asymmetric and $l$-periodic potentials $W_1$ and $W_2$ representing polarity and periodicity of the filament \cite{Astumian:1994cl,Magnasco:1994ee,pesk94,Prost:1994xh,Astumian:1996gd,juli97PhysRevLett.78.4510,juli97b,Parmeggiani:1999fq} (see Fig. \ref{cytoskeletonandcellmotility_fig15}a).
The dynamics of this system can be conveniently represented in terms of two coupled Fokker-Planck equations that describe the evolution of the probability density $P_i(x,t)$ of the motor to be in state $i=1,2$ at position $x$ at time $t$. Explicitly, we have:
\begin{eqnarray}
\partial_t P_1 + \partial_x J_1 &=& -\omega_1 P_1 + \omega_2 P_2,\nonumber\\
\partial_t P_2 + \partial_x J_2 &=& \omega_1 P_1 - \omega_2 P_2.
\end{eqnarray}
The currents $J_i$ ($i=1,2$) result from diffusion, interaction with the potentials $W_i$, and the external force $f_{\rm{ext}}$:
\begin{equation}
J_i=\mu_i\left[ -k_B T \partial_x P_i - P_i \partial_x W_i + P_i f_{\rm{ext}} \right ].
\end{equation}
The transition rates $\omega_i(x)$ ($i=1,2$) between the two states are driven out of equilibrium by ATP consumption, whose strength can be represented by a single parameter $\Omega$ using the following form:
\begin{equation}\label{Eq.BreakingDetailedBalance}
\omega_1(x)=\omega_2(x) \exp{\left[ \frac{W_1(x)-W_2(x)}{k_B T} \right]} + \Omega\,\Theta(x),
\end{equation}
where $\Theta(x)$ is a $l$-periodic function of integral one over one period. For $\Omega=0$, detailed balance is satisfied. Within this formalism, it has been shown that both spatial symmetry and detailed balance need to be broken for directed motion to occur, which has been quantified in terms of an effective potential $W_{\rm{eff}}$ \cite{Prost:1994xh}.

\subsection{Coupled motors and spontaneous oscillations}

Directly interesting for eukaryotic cellular motility driven by cilia and flagella, is the motion of motors with respect to their associated cytoskeletal filament when a collection of them is rigidly coupled. Such structures are typical of skeletal-muscle structures (where Myosin II motors associate into the so-called ``thick filaments''), or of the axonemal structure that drives oscillatory motions in cilia and flagella. Such systems have been studied using the two-state thermal-ratchet model \cite{Julicher:1995yj,juli97PhysRevLett.78.4510}, and a crossbridge model \cite{HUXLEY:1957ys,HUXLEY:1957rt,Hill:1974bh,Brokaw:1975bq,Vilfan:1998uq,Vilfan:1999fk}. Here we shall discuss essentially the case of an ensemble of motors that are rigidly linked to each other and walk collectively on a cytoskeletal filament whose interaction with the motors is described by the two-state model \cite{juli97b}.  In the case of randomly distributed motors, or motors distributed periodically with a period $q$ that is incommensurate with the filament period $l$, the probability density $P(\xi,t)$ of finding a particle at position $\xi=x$ mod $l$ in either state $i=1$ or 2, approaches the value $1/l$ in the case of an infinitely-large number of motors. In a mean-field approximation, equations of motion for the probability densities read:
\begin{eqnarray}
\partial_t P_1 + v\partial_{\xi} P_1 &=& -\omega_1 P_1 + \omega_2 P_2,\nonumber\\
\partial_t P_2 + v\partial_{\xi} P_2 &=& \omega_1 P_1 - \omega_2 P_2.
\end{eqnarray}
The force-velocity curve can then be computed using the fact that $f_{\rm{ext}}=\eta v - f$, where $f_{\rm{ext}}$ is the external force applied, $\eta$ is the friction coefficient per motor protein, and $f$ is the force per motor protein exerted by the potentials:
\begin{equation}\label{eq.f}
f=-\int_0^a d\xi\,(P_1\partial_{\xi}W_1+P_2\partial_{\xi}W_2).
\end{equation}
Expressing $P_2$ as $P_2=1/l-P_1$, and $P_1$ as a series expansion in powers of the velocity $v$, one finds a generic series expansion for the force-velocity curve $f_{\rm{ext}}$ as a function of $v$ in the steady state. As a function of the distance to thermal equilibrium  $\Omega$, controlled by ATP consumption by the motors, and which appears as a control parameter for the dynamics with a critical value $\Omega_c$, the curve $f_{\rm{ext}}(v)$ can be strictly monotonic for $\Omega \leq \Omega_c$, or present some multi-valuated regions for $\Omega > \Omega_c$, where two stable velocity regimes exist for a given external force. For symmetric potentials, the system is quiescent with $v=0$ at zero force for $\Omega \leq \Omega_c$, but present two possible opposite spontaneous velocities for $\Omega > \Omega_c$, a spontaneous symmetry breaking  that is characteristic of second-order phase transitions with characteristic mean-field exponents. Such a reversible spontaneous movement has been observed in a motility assay with NK11 proteins, a mutant of the kinesin protein Ncd that has lost its directionality \cite{Endow:2000la}. In addition, when the external force is varied, a hysteresis is found for $\Omega > \Omega_c$,  an experimental observation of which has been reported for a myosin II motility assay under near-stalling conditions induced by electric fields \cite{Riveline:1998db}. Numerical simulations of both situations have been performed using the two-state model with a finite number of motors and in the presence thermal noise \cite{Badoual:2002kq}.

\subsection{\label{SSe.AxonemalBeating}Axonemal beating}

The previous and related models have been used to describe the spontaneous oscillations that have been observed in skeletal-muscle myofibrils' oscillations or axonemal beating \cite{Okamura:1988qo,Yasuda:1996rp,Fujita:1998la}. In these cases, it has been proposed that the coupling of the motor backbone to a spring prevents spontaneous steady-state velocities to occur, but instead leads to spontaneous oscillations \cite{juli97PhysRevLett.78.4510} (see Fig. \ref{cytoskeletonandcellmotility_fig15}b). In the case of axonemal beating, of most relevance for eukaryotic cell swimming, the elastic force results from bending of the microtubules and leads to self-organization of the dynein motors. This collective behavior has been proposed to explain the bending waves of cilia and flagella \cite{mach58,Brokaw:1975bq} and analyzed in the framework of the two-state model \cite{Camalet:1999qy,Camalet:2000uq}. Close to the oscillatory instability, wave-patterns can be computed, whose frequencies and shapes depend on the filament length and boundary conditions, and which are in good agreement with observed flagellar beating patterns \cite{Vernon:2004wq}.

In the case of cilia however, beating patterns are typically assymetric, like it is at best exemplified in the case of the two cilia of the green alga \emph{Chlamydomonas} \cite{brayB}, an observation that cannot be accounted for by investigating beating patterns at the oscillatory instability only. Using the same underlying model, it has been proposed that this assymetry originates from the presence of transverse external flow that occurs as the organism is swimming \cite{Guirao:2007th}. In that case, the cilium tends to beat faster and quite straight in the direction of the flow, whereas it comes back slower and more curved against it, a beating pattern that evokes power and recovery strokes. Hydrodynamics has also been proposed to be responsible for dynamic coupling of adjacent cilia, which results in both spontaneous symmetry breaking and synchronization of their beating pattern \cite{Guirao:2007th}. This effect could be at the basis of the observed beating waves that propagate for example on the surface of \emph{paramecia}\footnote{An illustration of a paramecium can be seen in Fig. \ref{cytoskeletonandcellmotility_fig1}, center.} as they swim, which originate from a constant phase difference in the beating of the adjacent cilia, and which have been called \emph{metachronal waves} \cite{brayB,Gueron:1997fb,Guirao:2007th}. This could also underly the breaking of symmetry that occurs in mammalian development during gastrulation, and which is responsible for left-right asymmetry. In that case, it has been shown that beating of cilia located in a transiently-formed epithelial chamber known as the \emph{node}, create a directional flow which transports signaling molecules preferentially to one side \cite{brayB,Ibanez-Tallon:2003bd,McGrath:2003zt}. There, beating patterns are unusual in that cilia swirl in vortical fashion rather than beat \cite{Okada:2005hq}, and hydrodynamic-driven synchronization of these three-dimensional beating patterns has also been studied \cite{Vilfan:2006fu}.

\section{\label{Se.ActiveGels}Putting It Together: Active Polymer Solutions}

The last part of this review is devoted to the presentation of some generic descriptions of the cell cytoskeleton, when considered as a network of long protein filaments that are cross-linked by a variety of smaller proteins. As already discussed, filamentous proteins that are involved in the cell-cytoskeleton dynamics are mostly F-actin and microtubules (made of G-actin and tubulin monomers), with which interact cross-linkers that can be either passive and stationary (such as $\alpha$-actinin), or active and mobile, consisting then of clusters of molecular motors (mostly myosin and kinesin motors) (see Section \ref{Se.Cytoskeleton}). To model these systems, different complementary approaches have been developed, namely computer simulations \cite{Nedelec:1997rm,Surrey:2001yg,Nedelec:2001zl}, and analytical descriptions that can be roughly divided into three categories, namely microscopic, mesoscopic, and macroscopic or phenomenological hydrodynamic descriptions. The first analytical approaches that have been developed correspond to mesoscopic descriptions. There, starting from a microscopic description of the filaments, the effect of active cross-linkers is described via motor-induced relative velocities of paired filaments, where the form of such velocities is inferred from general symmetry considerations \cite{Nakazawa:1996fj,seki98,Kruse:2000ys,Kruse:2001fr,Kruse:2003th}. Microscopic approaches start from what is known about the properties of the different molecular players involved and their interactions, and aim to build large-scaled coarse-grained theories from statistical physics' principles \cite{Liverpool:2001fv,Liverpool:2005rt,Ahmadi:2005hl,Ahmadi:2006qd,Kruse:2006sf,Liverpool:2006rc,Aranson:2005rw,Aranson:2006dn}. Finally, macroscopic hydrodynamic approaches have adopted a more phenomenological point of view: they harness the generic symmetry and dynamical properties of the players involved, to derive directly effective continuous theories in terms of a few coarse-grained fields \cite{Toner:1998ys,Ramaswamy:2000db,Lee:2001gd,Aditi-Simha:2002wd,Hatwalne:2004cj,Sankararaman:2004om,Kruse:2004ek,Kruse:2005rt,Zumdieck:2005la,Voituriez:2006lh,Kruse:2006la,Julicher:2007dq,Joanny:2007vn}. Recently, attempts have been made to bridge microscopic to macroscopic models, and compare what results in being similar and different in the two types of approaches \cite{Liverpool:2003lq,Ahmadi:2006qd,Liverpool:2006rc}.

\begin{figure}[h]
\scalebox{0.37}{
\includegraphics{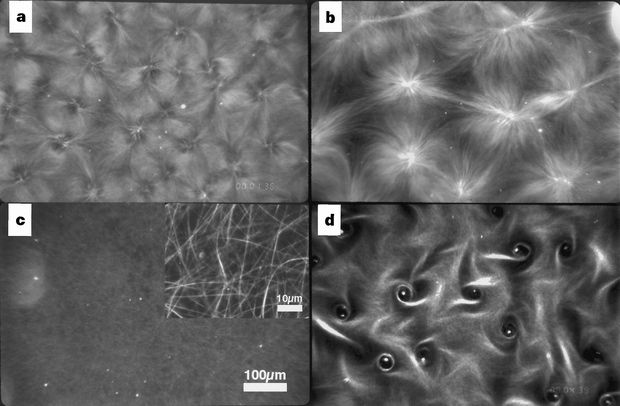}
}
\caption[cytoskeletonandcellmotility_fig16]
{\label{cytoskeletonandcellmotility_fig16}
Different large-scale patterns formed through self-organization of microtubules and kinesin motors as reported in ref. \cite{Nedelec:1997rm}. Initially uniform mixtures of proteins heated to $37\,^{\circ}\mathrm{C}$ displayed different patterns after 7 min of self-organization. Patterns are shown at equal magnification; the samples differ in kinesin concentration. a, A lattice of asters and vortices obtained at 25 $\rm{g}.\rm{ml}^{-1}$ kinesin concentration. b, An irregular lattice of asters obtained at 37.5 $\rm{g}.\rm{ml}^{-1}$ kinesin concentration. c, Microtubules form bundles at 50 $\rm{g}.\rm{ml}^{-1}$ kinesin concentration (scale bar, 100 $\mu$m); insert: at higher magnification (scale bar, 10 $\mu$m). d, A lattice of vortices obtained at a kinesin concentration smaller than 15 $\rm{g}.\rm{ml}^{-1}$. Source: courtesy of Fran\c cois N\'ed\'elec; reprinted from ref. \cite{Nedelec:1997rm} with permission from Nature Publishing Group.}
\end{figure}

Interests for describing the cytoskeleton as an ensemble of filamentous polymers actively connected by cross-linkers have come to the scene since self-organizations of motor-filament mixtures were observed experimentally \cite{Urrutia:1991db,Nedelec:1997rm,Surrey:2001yg}. Among these, complex patterns that include asters, vortices, spirals and connected poles or networks have been observed in confined quasi-two-dimensional systems in \emph{in vitro} experiments \cite{Nedelec:1997rm,Surrey:2001yg} (see Fig. \ref{cytoskeletonandcellmotility_fig16}).
Patterns where shown to be selected in a way that is dependent on motor and ATP concentrations, and numerical simulations based on microscopic models of rigid rods connected by active elements have shown to be capable of reproducing the experimental results \cite{Surrey:2001yg}. Further experiments were performed on systems that resemble more closely a living cell, and which, while being simplified versions of it, still exhibit some of its behaviors. Along these lines, formations of bipolar spindles that do not contain any microtubule-organizing center were observed using cell extracts \cite{Hyman:1996cs}, and cell fragments that contain only the actin cortex where found to self-propagate on a substrate, with coexistence of locomoting and stationary states \cite{Euteneuer:1984pi,Verkhovsky:1999hb}.

\subsection{Mesoscopic approaches}

Theoretical modeling of the cell cytoskeleton have benefitted from the knowledge accumulated in equilibrium statistical physics of polymer solutions and liquid crystals \cite{deGennesProstB}. However, the cell cytoskeleton is an active medium for which new analysis techniques needed to be developed in order to describe, for example, its ability to actively self-organize, exert forces and create motion. First, theoretical models have aimed to describe pattern formations in systems of actively-driven rigid filaments in one-dimensional geometries \cite{Kruse:2000ys}. Such configurations are represented \emph{in vivo} for example by stress fibers that are important for cell-force generation, contractile rings that form during cytokinesis, or the formation of filopodia for forward protrusion during amoeboid motility. There, dynamical equations that govern individual filaments were introduced using ``mesoscopic'' mean-field models, where the relative sliding of paired polar filaments is described by an effective relative velocity that is induced by many individual events of motor activity. General constraints on these relative velocity fields are imposed by symmetry considerations that rely on the orientational polarity of the filaments (see, e.g., \cite{Kruse:2000ys}). In such systems, polarity sorting \cite{Nakazawa:1996fj}, contraction \cite{seki98,Kruse:2000ys}, as well a propagating waves \cite{Kruse:2001fr} emerge from the models. Interestingly, it has been shown that relative velocity of filaments of the same orientation is important for contraction to occur \cite{Kruse:2000ys}, a phenomenon that has been suggested to rely on motor-density inhomogeneities along the filament that create inhomogeneous filament interactions along their lengths \cite{Kruse:2002rz}. More generally, the whole bifurcation diagram of this generic one-dimensional model has been established, and motor-distribution dynamics has been introduced that lead to contractile states with the generation of contractile forces, of most relevance for stress fibers' as well as contractile rings' dynamics \cite{Kruse:2003th}. Interestingly, the simplest version of these models, when only one possible polymer orientation is considered, has been mapped to hopping models that describe driven-diffusive systems \cite{Mohanty:2007kx}. In the absence of active cross-linkers, the model reduces to a class of hopping models known as the Zero Range Process (ZRP), for which exact analytical solutions of the steady state as well as one-dimensional phase transitions have been described \cite{Hanney:2004oq}. In the generic case however, the dynamics defines a new class of driven-diffusive systems, which can still be mapped in some cases to the ZRP analytic solution, even though with a different criterium for condensation to occur \cite{Mohanty:2007kx}.

\subsection{Microscopic approaches}

Microscopic approaches to describing the cell cytoskeleton dynamical behavior model explicitly all the considered different processes and interactions that occur between the different molecular players, and aim to derive effective dynamical equations for the different density fields that enter the description, by coarse-graining the microscopic dynamics. Most studies that have done so model the motor-filament system as an ensemble of rigid rods of fixed lengths, which interact via point-like cross-linkers that can induce relative sliding as well as rotational motions \cite{Liverpool:2005rt,Ahmadi:2005hl,Ahmadi:2006qd,Kruse:2006sf,Liverpool:2006rc}. Exceptions to this rule are theoretical descriptions of the mechanical response of active-filament solutions to high frequency stimuli \cite{Liverpool:2001fv,Liverpool:2003eu}. There, anomalous fluctuations occur that are dominated by the bending modes of the filaments in combination with the activity of the cross-linkers. Inspired by polymer physics at thermodynamic equilibrium, excluded-volume interactions as well as entanglements are taken into account in the description. In particular, the system exhibits accelerated relaxation at long times due to directed reptation that relies on active phenomena.

Attempts at deriving the motor-mediated interaction between filaments from microscopic descriptions have been performed in \cite{Aranson:2005rw,Aranson:2006dn,Liverpool:2005rt,Ahmadi:2005hl,Ahmadi:2006qd}. See also the review \cite{Liverpool:2006qr}. In \cite{Aranson:2005rw,Aranson:2006dn}, a generalization of the Maxwell model of binary collisions in a gas is used to describe the dynamics of polar rods whose inelastic and anisotropic interactions reflect the presence of active crosslinkers. Orientational instabilities lead to bundling as well as the formation of asters and vortices patterns. In \cite{Liverpool:2005rt,Ahmadi:2005hl,Ahmadi:2006qd}, filaments are described as rigid rods of fixed length, and hydrodynamics is obtained by coarse-graining the Smoluchowsky equation for rods in solution, coupled via excluded-volume and motor-mediated interactions. There are two main motor-mediated mechanisms for force exchange among filaments. First, active crosslinkers induce bundling of filaments, building up density inhomogeneities. Second, they induce filament sorting as a function of their polarization state. As a result, phase diagrams are derived that show instabilities of the homogeneous states at high filaments' and crosslinkers' densities. In particular, all homogeneous states are rendered unstable by the same mechanism of filament bundling, a fact reminiscent of the effect described in \cite{Kruse:2000ys} where the interaction between filaments of the same orientation has been shown to be important for contraction to occur. Interestingly, the broken directional symmetry of the polarized state yields an effective drift velocity that describes filament advection. This convective-type term describes a genuine out-of-equilibrium contribution that is structurally not present in phenomenological descriptions based on systematic linear expansions close to thermodynamic equilibrium (see below). Such a term is reminiscent of the one introduced in earlier studies of self-propelled nematic particles \cite{Toner:1998ys,Aditi-Simha:2002wd,Ramaswamy:2003fr}, as well as of the explicit flow of the solvent taken into account in \cite{Kruse:2004ek,Kruse:2005rt,Voituriez:2006lh}. Other effects of higher-order nonlinear terms have also been discussed in \cite{Ziebert:2004db,Ziebert:2005qd}, where pattern selection between stripe patterns and periodic asters occurs via nonlinear interactions.

\subsection{\label{Sse.Gels}Macroscopic phenomenological approaches: The active gels}

The third category of approaches that have been developed to try to understand the dynamical behavior of the cell cytoskeleton as a whole, are effective phenomenological theories which rely on the hypothesis that large length- and time-scales behaviors of the cytoskeleton are largely independent of the microscopic details that underly its dynamics, but depend instead only on a few macroscopic fields that capture the relevant behavior. Sufficiently close to thermodynamic equilibrium, these relevant fields describe the \emph{hydrodynamic modes} (or \emph{slow modes}) of the dynamics, namely the modes whose relaxation rates go to zero at long wavelengths. As for equilibrium systems close to a critical point, such hydrodynamic modes correspond to the \emph{conserved densities} on the one hand, and the \emph{order parameters} that break continuous symmetries on the other hand \cite{HOHENBERG:1977ly,deGrootMazurB}. To write generic theories for the dynamics of these hydrodynamic modes, standard approaches consist in writing systematic expansions in the different couplings that are allowed by the symmetry properties of the system. In the vicinity of a critical point, which occurs generally when a continuous symmetry is spontaneously broken after a second-order phase transition has been traversed, the concept of renormalization group has given a theoretical framework to identify universality classes: starting from the full nonlinear expansions of the underlying stochastic dynamics, only a few relevant parameters matter for the asymptotic scaling laws that occur at the transition \cite{zinnB,HOHENBERG:1977ly,Fisher:1998gf}. Even though originally developed to study equilibrium critical points, the renormalization-group concept has allowed for the characterization of some out-of-equilibrium universality classes (see e.g. \cite{Frey:1994qe,Tauber:2005ul,Risler:2004ao} and the review \cite{Hinrichsen:2006lq}).

Away from such remarkable points however, any term allowed by symmetry in a systematic expansion is \emph{a priori} relevant. The standard approach for systems close to thermodynamic equilibrium consists in writing generalized thermodynamic forces and fluxes that are related to each other by linear-response theory. Constraints on the generic coupling constants to linear order emerge from the spatio-temporal symmetries of the system, and correspond to the Onsager relations and the Curie principle \cite{deGrootMazurB}. Inspired by the dynamics of liquid crystals \cite{mart72,deGennesProstB}, a hydrodynamic theory has been developed that describes the cytoskeleton as a visoelastic polar gel, driven out of equilibrium by a source of chemical energy \cite{Kruse:2004ek,Kruse:2005rt,Zumdieck:2005la,Voituriez:2006lh,Kruse:2006la} (this work has been reviewed in \cite{Julicher:2007dq}). Among other applications, such or similar approaches have been applied to the description of pattern formation in motor-microtubule mixtures \cite{Lee:2001gd,Sankararaman:2004om}, as well as the collective dynamics of self-propelled particles \cite{Toner:1998ys,Aditi-Simha:2002wd,Hatwalne:2004cj,Dombrowski:2004rw}.

Originally, this hydrodynamic theory has been presented as a generic theory for active viscoelastic materials made of polar filaments, referred to as \emph{active polar gels} \cite{Kruse:2005rt}. Within the framework of the previously-described general formalism, here applied to the cytoskeleton, conserved quantities are the different number densities that enter the dynamics, namely the number densities of subunits in the gel, of free monomers, and of respectively bound and unbound motors to the filaments. To these must be added the solvent density and the total mechanical momentum. Source terms in the conservation equations correspond to polymerization and depolymerization of cytoskeleton filaments, attachment and detachment of motor proteins to the filaments, and the potential presence of an external force. Order parameters correspond to orientational order parameters that originate from the polarity of the filaments. Namely, they correspond to momenta of the local polarization vector of individual filaments $\bold{u}$, and most often only the first momentum $\bold{p}=\langle \bold{u} \rangle$ is considered that represents the locally-averaged polarity in the gel\footnote{The next momentum $q_{\alpha\beta}=\langle u_{\alpha}u_{\beta} - d^{-1}\,\bold{p}^2\,\delta_{\alpha\beta}\rangle$, where $d$ is the dimension of space, is a symmetric traceless tensor of order two that corresponds to nematic order.}. To these must be added a crucial parameter that drives the system out of equilibrium, and which originates from the actively-maintained source of chemical energy in the cell, corresponding to out-of-equilibrium concentrations of ATP versus ADP and $\rm{P}_{\rm{i}}$. This parameter $\Delta\mu$ is expressed as the difference in chemical potentials of ATP versus ADP plus $\rm{P}_{\rm{i}}$: $\Delta\mu=\mu_{\rm{ATP}}-(\mu_{\rm{ADP}}+\mu_{\rm{P_{\rm{i}}}})$. After identification of the different conjugated generalized fluxes and forces, that are split into dissipative and reactive parts as a function of their properties under time-reversal symmetry, the constitutive equations that specify the dynamics are written in terms of a generalized Maxwell model, which describes the viscoelastic dynamical properties of the gel. Under its simplest form and for nonpolar viscoelastic gels, the Maxwell model writes
\begin{equation}
\left( 1+\tau\frac{D}{D t}\right)\sigma'_{\alpha\beta}=2\eta\left(v_{\alpha\beta}-\frac{1}{d}\delta_{\alpha\beta}v_{\gamma\gamma}\right)+\bar{\eta}\delta_{\alpha\beta}v_{\gamma\gamma}
\end{equation}
in $d$ dimensions. Here $v_{\alpha\beta}$ and $\sigma'_{\alpha\beta}$ are the symmetric parts respectively of the velocity-gradient tensor $\partial_{\alpha}v_{\beta}$ and the viscous stress tensor, $\eta$ and $\bar{\eta}$ are respectively the shear and bulk viscosities, and $\tau=E/\eta$ is the viscoelastic relaxation time that is related to the Young elastic modulus $E$ and to the shear viscosity $\eta$. This relaxation time describes the crossover between an elastic behavior at short times that resembles that of a solid gel and a viscous behavior at long times that resembles that of a fluid\footnote{Note that only one relaxation time is assumed to characterize the system, as some experiments suggest that a power-law distribution of relaxation times is better suited to describe cytoskeleton dynamics, potentially because of some scale-invariant dynamical properties in the system \cite{Fabry:2001ek,Balland:2006oa}.}. Finally, $D/D t$ represents a convective corotational derivative that takes into account invariance with respect to translations and rotations in the system. In the general framework of active polar gels close to thermodynamic equilibrium, generic linear couplings are added to this model following the general procedure described above. Extensive presentations of the complete formalism can be found in refs. \cite{Kruse:2005rt,Julicher:2007dq}, which also include discussions about its limitations and some of its possible extensions. In particular, in ref. \cite{Julicher:2007dq}, extensions that aim to include the contributions of rotational viscoelasticity, of some nonlinear couplings and of passive as well as active sources of noise are briefly discussed. The main developed extension so far concerns the generalizations of this formalism to multi-component active gels that are now being developed \cite{Joanny:2007vn,live08}. These allow in particular to take into account the possible permeation of the cytosol through the cytoskeletal gel, which affects force balance in the system, and which might be of importance for cell motility.

Despite its recent development, this generic theory of active polar gels has been applied to the description of some systems that are of particular relevance for cell motility or experimental situations observed \emph{in vitro}. Its first application was the study of topological defects in the polarity field of the gel that lead to the formation of patterns such as asters, vortices and spirals \cite{Kruse:2004ek}. As a function of two dimensionless parameters representing the relative strength of the coupling to the chemical potential $\Delta\mu$ and to the bend and splay moduli of the polar gel, a phase diagram was derived where vortices and asters give rise to rotating spirals via dynamic instabilities. These relate to the spatial patterns that have been observed \emph{in vitro} \cite{Nedelec:1997rm,Surrey:2001yg}, as well as to the creation of spontaneous motion, of most relevance for cell motility. In a different geometry, namely a cylinder of finite diameter and length, the formalism has been applied to establish a phase diagram of ring formation that contains phases of one or multiple rings, and which can be quiescent or oscillating \cite{Zumdieck:2005la}. This is relevant for understanding the formation and localization of cortical rings that form prior to cytokinesis and for which double-ring formation has been observed with certain plant cells \cite{Granger:2001lr}. To understand the generation of active flows that might be of relevance for cell crawling, a generic phase diagram has been derived for a two-dimensional active polar film that is compressible \cite{Voituriez:2006lh}. Compressibility here might refer to different thicknesses in a three-dimensional incompressible gel that is described in two dimensions after integration of the density fields over its thickness. Within this framework, density fluctuations couple generically to polarity splay, and different topological phases of the gel-polarity organization are found that could correspond to some of the previously-observed patterns in the experimental literature. Finally, the description of spontaneous movements of thin layers of active gels has been applied to the study of cell locomotion on a solid substrate that occurs via the protrusion of the actin-filled lamellipodium at the leading edge of the cell \cite{Kruse:2006la}. Reducing the lamellipodium description to a two-dimensional gel protruding in one dimension, and with a spatially-dependent thickness, the steady-state thickness profile as well as the flow and force fields have been computed. One particularly striking aspect of cell crawling that is described by this formalism is the presence of a retrograde flow of the gel as the cell is crawling. This aspect has been quantified in earlier experiments performed on fish epidermal keratocytes \cite{Vallotton:2005fy}. It has been shown that while the cell is crawling, treadmilling of actin filaments happens faster than global motion of the cell, such that the actin cortex is moving rearward with respect to the substrate, in a direction opposite to the movement of the cell \cite{Lin:1996cr,Coussen:2002sp,Jurado:2005wd}. Similar questions have been addressed using different theoretical frameworks in \cite{Alt:1999kk,Rubinstein:2005rr,Kozlov:2007fk}.

\subsection{\label{Sse.Comparisons}Comparisons of the different approaches to describing active polymer solutions}

With these different ways of approaching the description of the dynamics of the cell cytoskeleton as a whole, a natural question is to ask to what extent these different approaches are similar and different, and which aspects of the cytoskeleton or cell behavior can be or not described by each of the theories. For answering these questions, connections between the different approaches have been made, first between mesoscopic and hydrodynamic descriptions \cite{Liverpool:2003lq,Kruse:2003yq}. In \cite{Liverpool:2003lq}, a generalization of the mesoscopic model introduced in \cite{Kruse:2000ys,Kruse:2001fr} is developed to obtain a set of continuum equations in unconfined geometries. A phase-diagram is derived that results from the stability analysis of the homogeneous state of actively cross-linked polymers, taking into account excluded-volume interactions and estimates of entanglement in two and three dimensions. It is found that an instability occurs as the bundling rate between filaments of the same orientation is increased, which at low filament density happens first via a density-fluctuation instability, and at high filament density via an orientational-fluctuation instability. In the presence of a finite sorting rate between filaments of different orientations, propagating modes appear that reflect oscillatory behavior. In \cite{Kruse:2003yq}, the continuum theory is related to nonlocal descriptions of filament-motor systems, since filaments can transmit stresses over finite distances. The effective parameters of the continuum theory are recovered from the previously-published mesoscopic description \cite{Kruse:2000ys}, even though with missing coefficients that are thought to correspond to microsocpic multi-particle interactions, not described in \cite{Kruse:2000ys}. Effects of polymerization-depolymarization dynamics via effective source and sink terms in the local filament densities are also discussed (see also \cite{Joanny:2007vn,live08}) - like it is the case in the effective macroscopic theories - as well as the role of polarity. In particular, it is found that nonpolar arrangements of filaments do not exhibit oscillatory instabilities and propagating modes, which might be of relevance for muscle sarcomeric structures. As seen previously, in these systems, spontaneous oscillations that have been observed correspond more to oscillatory instabilities of rigidly-coupled collective motors than to solitary-wave solutions, as they are found in systems of active polar filaments.

Microscopic theories present the advantage of being able in principle to give rise to a full description of a given system with arbitrary precision and specificity, and to take into account the nonlinear effects that are of direct relevance for the system's behavior. However, they rely on the microscopic knowledge that one has on the system under consideration, and are therefore limited by the available information on the different agents. In addition, they end up with effective descriptions that are model-dependent, in that the different parameters of the so-obtained theory, which describe its physical behavior, depend on the interactions that are taken into account at the microscopic level. Also, an important aspect of active cytoskeleton dynamics that is usually not described in such microscopic approaches is the very important phenomenon of treadmilling that relies on polymerization-depolymerization dynamics of cytoskeletal polymers, and which we have seen to be of crucial importance for some mechanisms of cellular motility such as \emph{Listeria} propulsion or nematode-sperm-cell locomotion (see Section \ref{Se.Filaments}). However, despite the absence of these effects, which are taken into account effectively in macroscopic hydrodynamic descriptions, such microscopic approaches allow for the derivation of the forces exchanged between the motors and the filaments from microscopic knowledge, while they appear as effective parameters of unknown explicit origin in effective macroscopic descriptions. Thereby, questions can be addressed that concern the role played by the specific physical properties of motor-filament interactions at the microscopic level in controlling the system behavior on large scales. Indeed, the richness of the observed self-organized structures raises the question of how much is generic, and how much is specific in cytoskeleton behavior. For example, experiments have shown that very different self-organizing structures occur with processive as opposed to non-processive motor proteins: at high motor concentrations, microtubule-kinesin mixtures self-organize in a variety of spatial patterns \cite{Nedelec:1997rm,Surrey:2001yg}, as homogeneous states are more robust with acto-myosin systems \cite{Humphrey:2002mw}, an effect that can be thought of as the influence of motor processivity on the dynamical large-scale parameters \cite{Liverpool:2003lq}.

\section{\label{Se.Future}Extensions and Future Directions}

Cell motility is a complex and integrated process that relies on self-organization of the cytoskeleton, carefully and precisely orchestrated by the cell with the help of numerous different types of molecular players. If one includes the subcellular movements that are responsible for intracellular traffic and material exchange between the inner parts and external parts of the cell, cell-motility mechanisms are found to ground the activity of all life forms on earth. When looked under the microscope, motility mechanisms and structural changes of the diverse cell types appear so vast and various that a comprehensive understanding of their underlying mechanisms seems to be an overwhelming challenge. However, as we have seen from the literature covered in the present article, our understanding of cell motility has tremendously progressed over the past two decades. On the one hand, complexity has even further emerged, since the biochemical characterization of the molecular players involved has revealed that at least hundreds of different protein types participate in the structural and dynamical organization of the cell cytoskeleton. On the other hand, despite the existence of such very complex regulation processes that rely on the integrated interplay of the whole set of different molecular players, the characterization of the cell cytoskeleton has revealed that its main structures and functions are due to just a few types of key proteins, namely three types of biopolymers and three superfamilies of molecular motors. Even more striking is the evolutionary conservation of the main molecular players involved in building the cytoskeleton dynamical filaments, both within the eukaryotic domain of life on the direct sequence point of view, and even across the three domains of life when structural and functional properties are considered. These striking observations indicate that the different underlying mechanisms of cell motility all rely on generic principles that can be understood on a biophysical point of view. In addition, further help from micro-manipulation and fluorescence-microscopy techniques, as well as the development of simplified systems based on gene-expression control and bio-mimetic artificial systems, has enabled the experimental biophysical investigation of the different specific aspects of the processes at play.

The theoretical analyses reviewed in this article have shown that central concepts that underly the cytoskeleton dynamics are self-organization and dynamic instabilities, here grounded on out-of-equilibrium nonlinear dynamics', thermodynamics' and statistical physics' principles. Such concepts are at the basis of all the theoretical approaches that have been developed to understand the mechanisms of diverse phenomena such as polymerization-depolymerization force and movement generation, molecular motors' individual behaviors and collective phenomena, as well as the generic behaviors of active-polymer solutions which lead to a description of the cytoskeleton dynamics as a whole. On all of these topics, microscopic as well coarse-grained effective macroscopic approaches have been developed. As already discussed, they both have their advantages, powers and limitations, and represent important complementary steps in the ultimate goal of an integrated description of the universal principles that underly cell motility.

As we have seen in this article, our understanding of cell motility and cell cytoskeleton dynamics has grandly benefitted from the interplay between experiments and modeling, each for its own reasons guiding the other in its directions of investigation. To further understand the integrated processes at play 
in cell motility, such fruitful interactions will certainly be further required and developed. On the theoretical point of view, bridges between understanding simplified systems or some particular aspects of cell motility and the phenomenon of cell motility as a whole at the global cellular level, have already started being investigated, but further developments of these two different ways of approaching the cytoskeleton dynamics as well as understanding the links that ultimately relate them are required. Another important aspect whose understanding represents a challenge is the potentially crucial role of noise that has been so far most of the time absent from the macroscopic effective theoretical approaches. Indeed, noise in nonlinear dynamical systems is known to potentially have important constructive effects, whose main representatives are stochastic resonance, coherence resonance and noise-induced transitions, as well as the extensive gallery of different spatially-extended phenomena such as array-enhanced stochastic and coherence resonance, or noise-enhanced synchronization of nonlinear oscillators (see e.g. \cite{pikoB}). Such phenomena have already been recognized to play an important role in some biological cytoskeleton-based pattern formations (see e.g. \cite{Howard:2003lr}), and could play a crucial role in driving other cytoskeletal self-organization phenomena, especially close to dynamical instabilities, where the effect of noise is highest. Finally, having at hand the underlying biochemical and biophysical mechanisms of cell forces and motility, a great challenge is to understand self-organization at yet larger scales, namely in animal tissues, where collections of cells present integrated coherent behaviors that drive diverse key processes such as morphogenesis, wound healing, immune response, tumor development and metastases formations. There, the same scheme involving ``microscopic'' as well as effective ``macroscopic'' approaches can certainly play an equally major role, ``microscopic'' approaches then potentially integrating the whole knowledge acquired at the level of a single cell, and partially reviewed in this article.

\section*{\label{Se.Acknowledgments}\large{Acknowledgments}}

To write this article, I grandly benefited from Karsten Kruse's habilitation thesis, which constituted a very good starting point as an extensive review of the existing biophysical literature on the cytoskeleton. I acknowledge C\'ecile Sykes and Jean-Fran\c cois Joanny for discussions, careful reading of the manuscript and constructive criticisms and suggestions. I also acknowledge Andrew Callan-Jones for pointing to me important references.

\section{\label{Se.Biblio}Bibliography}

\section*{\large{Books and Reviews}}

\begin{description}
\item {\bf Books}
\begin{description}
\item Alberts B {\it et~al.} (2002) {\em Molecular biology of the cell}, edited by S. Gibbs (Garland, New York), Chap. 16 and 18.
\item Bray D (2000) {\em Cell movements}, edited by M. Day (Garland, New York).
\item Howard J (2001) {\em Mechanics of Motor Proteins and the Cytoskeleton} (Sinauer Associates, Inc., Sunderland, Massachusetts).
\end{description}
\item {\bf Collections}
\begin{description}
\item Lenz P (ed) (2008) {\em Cell Motility (Biological and Medical Physics, Biomedical Engineering)}, ``Springer Science and Business Media'', LLC, New York).
\end{description}
\item {\bf Reviews}
\begin{description}
\item Reviews of special interest and that cover the subjects treated in this article can be found in the following references: \cite{reim02,juli97b,Fletcher:2004th,Small:2002sw,Kaverina:2002bs,Plastino:2005yq,Huxley:2004lr,Pollard:2003lr,Ananthakrishnan:2007kx,Shih:2006dq,Erickson:2007rr,Julicher:2007dq,Liverpool:2006qr,Ridley:2003wj}.
\end{description}
\end{description}

\end{document}